\documentclass[a4paper,fleqn]{cas-sc}

\usepackage[numbers]{natbib}
\usepackage{mathtools}
\usepackage{datatool}
\usepackage{steinmetz}
\usepackage{setspace}

\makeatletter

\newtheorem{definition}{Definition}[section]
\newcommand{\alphah}{\hat{\alpha}}
\newcommand{\betah}{\hat{\beta}}
\newcommand{\gammab}{\boldsymbol{\gamma}}
\newcommand{\alphab}{\boldsymbol{\alpha}}
\newcommand{\phib}{\boldsymbol{\phi}}

\DTLnewdb{bibnotes}

\def\bibnote#1#2{%
  \DTLnewrow{bibnotes}
  \DTLnewdbentry{bibnotes}{mylabel}{#1}
  \DTLnewdbentry{bibnotes}{mynote}{#2}
}
\makeatletter
\patchcmd{\@lbibitem}%
  {\item[\hfil\NAT@anchor{#2}{\NAT@num}]}%
  {%
    \item[\hfil\NAT@anchor{#2}{\NAT@num}]%
    \DTLforeach[\DTLiseq{\mylabel}{#2}]{bibnotes}{\mylabel=mylabel,\mynote=mynote}{\textit{\mynote}}
  }{}{\message{^^JPatching failed^^J}}%

\makeatletter
\newcommand\blfootnote[1]{%
	\begingroup
	\renewcommand\thefootnote{}\footnote{#1}%
	\addtocounter{footnote}{-1}%
	\endgroup
}
\makeatother
\begin{document}
\let\WriteBookmarks\relax
\def\floatpagepagefraction{1}
\def\textpagefraction{.001}
\shorttitle{Fully Adaptive Time-Varying Wave-Shape Model}
\shortauthors{J Ruiz, G Schlotthauer, L Vignolo \& MA Colominas.}
\bibnote{stevenson2019dataset}{\emph{[dataset]}}
\bibnote{goldberger2000physiobank}{\emph{[dataset]}}

\title [mode = title]{Fully Adaptive Time-Varying Wave-Shape Model: Applications in Biomedical Signal Processing}

\author[1]{Joaquin Ruiz}[type=editor,
                        auid=000,bioid=1,
                        prefix=,
                        role=,
                        orcid=0000-0002-0201-2190]
\cormark[1]
\ead{joaquin.ruiz@uner.edu.ar}

\credit{Conceptualization, Methodology, Software, Visualization, Writing - Original Draft}

\address[1]{Institute for Research and Development in Bioengineering and Bioinformatics (IBB), CONICET-UNER, Ruta Prov. 11 Km. 10, E3100, Oro Verde, Entre R\'ios, Argentina}

\author[1]{Gastón Schlotthauer}[
                        prefix=,
                        role=,
                        orcid=0000-0002-2692-3255]
\ead{gaston.schlotthauer@uner.edu.ar}

\credit{Conceptualization, Methodology, Writing - Review & Editing, Funding Acquisition}

\author[2]{Leandro Vignolo}[
                        prefix=,
                        role=,
                        orcid= 	0000-0002-8176-8235]
\ead{ldvignolo@sinc.unl.edu.ar}

\credit{Conceptualization, Methodology, Writing - Review & Editing}

\address[2]{Research Institute for Signals, Systems and Computational Intelligence-Sinc(i). CONICET. Ciudad Universitaria UNL, Ruta Nac. No 168, km. 472.4, FICH, 4th floor (3000) Santa Fe, Argentina}
\author[1]{Marcelo A. Colominas}[%
   role=,
   suffix=,
   orcid=0000-0003-4418-7527,
   ]
\ead{macolominas@conicet.gov.ar}

\credit{Conceptualization, Methodology, Writing - Review & Editing, Funding Acquisition}

\cortext[cor1]{Corresponding author}

\begin{abstract}
    In this work, we propose a time-varying wave-shape extraction algorithm based on a modified version of the adaptive non-harmonic model for non-stationary signals. The model codifies the time-varying wave-shape information in the relative amplitude and phase of the harmonic components of the wave-shape. The algorithm was validated on both real and synthetic signals for the tasks of denoising, decomposition and adaptive segmentation. For the denoising task, both monocomponent and multicomponent synthetic signals were considered. In both cases, the proposed algorithm can accurately recover the time-varying wave-shape of non-stationary signals, even in the presence of high levels of noise, outperforming existing wave-shape estimation algorithms and denoising methods based on short-time Fourier transform thresholding. The denoising of an electroencephalograph signal was also performed, giving similar results. For decomposition, our proposal was able to recover the composing waveforms more accurately by considering the time variations from the harmonic amplitude functions when compared to existing methods. Finally, the algorithm was used for the adaptive segmentation of synthetic signals and an electrocardiograph of a patient undergoing ventricular fibrillation.
\end{abstract}

\begin{keywords}
	Oscillatory Signal Modeling \sep Time-Varying Wave-Shape Function \sep Biomedical Signal Processing \sep Signal Denoising \sep Signal Decomposition \sep Signal Segmentation
\end{keywords}
\maketitle

\doublespacing
\section{Introduction}
\blfootnote{Article accepted for publication on September 12, 2023. DOI: \url{https://doi.org/10.1016/j.sigpro.2023.109258}}\blfootnote{\copyright\ 2023. This manuscript version is made available under the CC-BY-NC-ND 4.0 license: https://creativecommons.org/licenses/by-nc-nd/4.0/}Time-varying signals carry important information about the systems that generate them. In many fields, including the biomedical setting, these systems are dynamic since their intrinsic properties change through time. These transitions can be abrupt when the system goes from a discrete state to the next; or gradual, where the internal properties of the system change gradually through time. These systems generate non-stationary signals with time-varying properties, which manifest the system's underlying dynamics. Heart-rate variations under stress and pitch changes in a singer’s voice during a performance are examples of dynamic changes manifesting as variations in the signal's instantaneous frequency (IF). Likewise, changes in the intensity of the system’s response are associated with variations in the instantaneous amplitude (IA). Additionally, the shape of the oscillations can change. The combination of these different sources of variability leads to very complex patterns in the time evolution of the signal.

\begin{figure}
    \centering
    \includegraphics[width=\columnwidth]{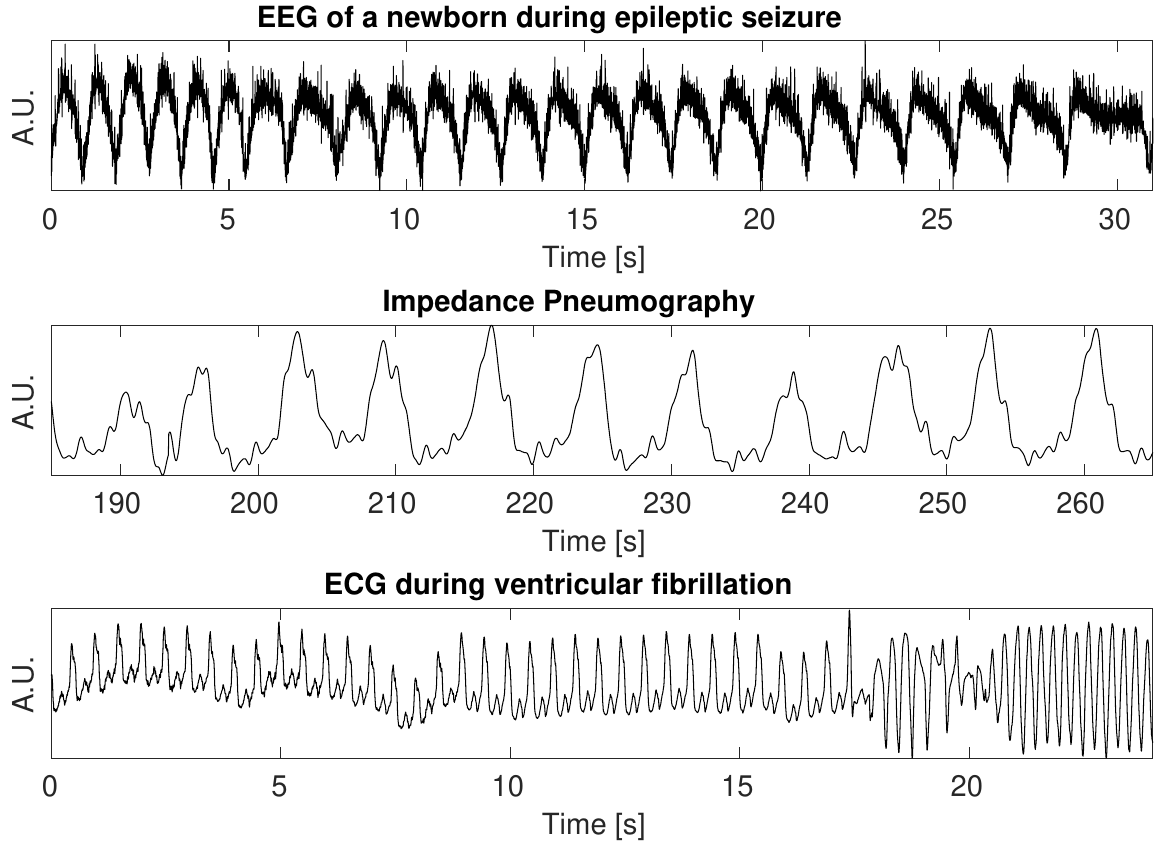}
    \caption{Examples of non-stationary biomedical signals with time-varying wave-shape. Top: EEG signal of a newborn experiencing an epileptic seizure event. Middle: Impedance pneumography composed of a respiratory and cardiac component. Bottom: ECG signal of a patient on the onset of a ventricular fibrillation event.}
    \label{fig:intro}
\end{figure}

Fig. \ref{fig:intro} shows some examples of real-world non-stationary signals, particularly biomedical signals. The first signal is an electroencephalography (EEG) signal of a newborn experiencing an epileptic seizure event. The recording is contaminated by muscle contraction noise, and the oscillatory pattern tends to change with time, while the ``cycle'' duration also changes. The second signal is an impedance pneumography (IP) signal, which is composed of a low-frequency respiratory component and a high-frequency cardiac component. The last signal is an electrocardiography (ECG) signal of a tachyarrhythmic patient on the onset of a ventricular fibrillation event, where the waveform clearly changes abruptly towards the end of the recording. We can see in these signals the variability in the shape of the oscillations due to the underlying dynamics of the systems that generate them, and we recall here that this temporal variation of the oscillatory pattern is an important aspect that needs to be taken into account.

Various challenges or tasks arise when dealing with non-stationary signals like the ones mentioned previously. Signal \emph{denoising} is a processing task that involves the removal of interferences and artifacts from the signal to enhance its relevant information. Tools like Empirical Mode Decomposition (EMD), Variational Mode Decomposition (VMD) and wavelet thresholding have been used for this purpose~\cite{yang2015emd,liu2016variational,bayer2019iterative,poornachandra2008wavelet}. On the other hand, signal \emph{decomposition} implies the separation of a (multicomponent) signal into two or more monocomponent signals. Algorithms for this task include resonant filter design, time-varying filtering using EMD, intrinsic chirp decomposition and non-local median filters~\cite{su2017extract, chen2017intrinsic, selesnick2011resonance, li2017time}. Moreover, signal \emph{segmentation} seeks to identify and separate distinct portions of the signals that are associated with different states or events that arise during registration. Tools like energy optimization and statistical analysis~\cite{popescu2014signal,mahmoodi2005signal} have been successfully applied for signal segmentation. These are just some examples of many processing tasks regularly performed on non-stationary signals.

More over, time-frequency (TF) analysis tools have been used extensively on non-stationary signals for the tasks mentioned. These tools include the synchrosqueezing transform (SST)~\cite{daubechies2011synchrosqueezed,thakur2011synchrosqueezing}, and its variants like the demodulation SST~\cite{wang2013matching,jiang2017instantaneous} and the adaptive SST~\cite{li2020aadaptive,li2020badaptive}. Recently, the chirplet transform has been proposed for component retrieval of signals with components overlapping in the time-frequency domain~\cite{chui2021time,li2022chirplet}. Although these methods have shown high performance for instantaneous frequency estimation and mode retrieval, their main limitation is that they consider each signal component as a simple oscillatory mode modulated in amplitude and frequency. In general, signals have more complex oscillatory patterns, and the morphology of these oscillations carries relevant information about the underlying processes.

Machine and deep learning tools have been extensively applied to signal processing problems, reaching high performance based on task-specific metrics~\cite{wei2020gravitational,arsene2019deep,yu2019deep,sabour2022gated,kim2019convolutional,clarke2020deep,rasti2021aecg}. Nonetheless, these methods are generally task-specific and require costly and lengthy training processes. We believe that an adaptive and versatile signal analysis method would be desirable to develop a general framework for the different processing tasks that can arise in various technical fields. In this context, techniques based on wave-shape estimation emerge as an appealing tool since they can characterize the dynamics of the systems from the parameters of a mathematical model of the signals.

In general, a non-stationary signal can be represented as a superposition of simple oscillatory components, most commonly sine and cosine waves, subject to amplitude (AM) and frequency (FM) modulation. This basic model, known as the adaptive harmonic (AH) model, has been extensively used for signal analysis in various fields, including the study of biomedical systems.

An alternative model, proposed by HT Wu~\cite{wu2013instantaneous}, extends the previous model by considering more general oscillatory patterns for each component. This new model, known as the adaptive non-harmonic model (ANH), replaces the cosine waves for wave-shape functions (WSFs), which have non-sinusoidal oscillations. Nevertheless, the main limitation of this model is that the WSFs remain \emph{fixed} throughout time. Wave-shape variability is an important characteristic of biomedical signals, as the morphology of the oscillation varies considerably for different physiological and pathological situations. For example, medical conditions like hypoxia, arrhythmia or fibrillation are associated with changes in the waveform of the electrocardiograph (ECG). Current approaches in time-varying waveform analysis rely heavily on landmark detection in the time domain or spectral variability analysis in the frequency domain. These techniques depend on the signal under analysis and might not be suitable for pathological cases. Current research lacks a comprehensive framework for the analysis of signals with time-varying wave-shape. To address this issue, Lin et al.~\cite{lin2018wave} proposed a more general ANH model that considers a time-varying wave-shape through temporal variations in the amplitude and frequency of the harmonic components of the wave-shape. This model is postulated as a good candidate for analyzing biomedical signals with variable waveforms. However, its applicability remains limited, mainly because it is not clear how to characterize the variability of harmonic amplitudes and frequencies. 

A few approaches have been proposed for time-varying wave-shape estimation. Multiresolution Mode Decomposition (MMD)~\cite{yang2021multiresolution} estimates shape functions at different levels of detail to approximate the time-varying wave-shape using recursive diffeomorphism-based regression~\cite{xu2018recursive}. This method assumes that the wave-shape variability is codified in amplitude- and frequency-modulated shape functions. Alternatively, Shape-Adaptive Mode Decomposition (SAMD)~\cite{colominas2021decomposing} models the time-varying wave-shape as the result of a polynomic approximation for the harmonic phases, while leaving the relative harmonic amplitudes constant. The oscillatory wave-shape model~\cite{lin2021wave} considers a segmentation of the signal followed by a representation of the cycles in a high-dimensional Euclidean space. Next, dimensionality reduction techniques are applied to visualize the distribution of the cycles while preserving the geometric distribution from the high-dimensional data. Unlike previous methods, wave-shape oscillatory modeling does not recover the underlying wave-shape functions from the segmentation. Additionally, extracted cycles may overlap which results in information being shared between neighboring cycles.

In this work, we propose a novel algorithm for adaptive estimation of the time-varying wave-shape of oscillatory signals by modifying the ANH model and applying a curve-fitting algorithm. Unlike previous approaches, our model considers both the non-integer nature of ``harmonic'' phase functions and the non-proportionality of harmonic amplitude functions. This wave-shape extraction procedure also allows the characterization of wave-shape variability in AM-FM signals using a (relatively) low-dimension coefficient vector. We define a novel set of harmonic amplitude functions and use these functions (alongside non-integer multiples of the phase) to characterize the time-varying wave-shape. Additionally, we propose an adaptive procedure for extracting the wave-shape function parameters from the data that includes the automatic selection of all the algorithm hyperparameters. This algorithm was validated on both synthetic and real-world signals for various commonly encountered tasks in the field of signal processing.

The rest of this document is organized as follows. In Sec. \ref{sec:models}, the AH and ANH models are introduced. The proposed modification to the ANH model is presented, aiming to obtain a usable model for the characterization of signals with time-varying wave-shape. In Sec. \ref{sec:numerical}, the numerical implementation details for the model estimation algorithm are presented, with particular emphasis on how the amplitude of the harmonics is estimated. In Sec. \ref{sec:synth}, the performance of the wave-shape extraction algorithm is evaluated in three signal processing tasks: denoising of monocomponent and multicomponent signals, multicomponent signal decomposition and segmentation of signals with sharp transitions. Real-world applications for denoising, decomposition and adaptive segmentation of biomedical signals are proposed in Sec. \ref{sec:real}. Finally, our conclusions are discussed in Sec. \ref{sec:conclusions}.

\section{Non-stationary Oscillatory Signal Modeling}
\label{sec:models}
The first step in the study of non-stationary signals is the development of a phenomenological model that concisely describes the variation of signal characteristics. A classical approach is the adaptive harmonic (AH) model
\begin{equation}
	\label{eq:AH}
	x(t) = \sum_{k=1}^K A_k(t) \cos(2\pi\phi_k(t)),\ A_k(t),\ \phi'(t)>0,
\end{equation}

\noindent where $t\in\mathbb{R}^+_0$ indicates time. $A_k(t)$ and $\phi_k'(t)$ are respectively the IA and IF of the $k$-th component of $x(t)$. This model assumes \emph{slow-varying} conditions for both $A_k(t)$ and $\phi_k'(t)$:

\begin{itemize}
	\item[(C1)] $A_k(t)\in C^1(\mathbb{R}),\ |A'_k(t)|/A_k(t) < \epsilon_1$,
	\item[(C2)] $\phi_k(t)\in C^2(\mathbb{R}),\ |\phi_k''(t)| < \epsilon_2$, for all $t$ and for $k=1,2,\dots, K$ and $\|\phi_k''\| = M$.
\end{itemize}

When $K > 1$, we incorporate a separability condition $|\phi_k'(t) - \phi'_{k-1}(t)|> d$, for all $t$. This condition guarantees that the components do not overlap in the time-frequency plane. Note that we are assuming that the components are sorted according to their instantaneous frequencies in an increasing fashion. 

\subsection{Fixed Wave-Shape Model}
The previous model considers that each component of the signal oscillates like a cosine wave. In general, real-world signals have more complicated, non-sinusoidal oscillatory patterns. Biomedical signals are a prime example of this. Signals like the ECG, photoplethysmogram and airflow, among others, carry important physiological information that is encoded not only in the time-varying nature of its amplitude and frequency but also in the shape of the oscillations. In view of this, an alternative model was proposed by HT Wu~\cite{wu2013instantaneous}:

\begin{equation}
	\label{eq:ANH}
	x_{\textsc{fix}}(t) = \sum_{k=1}^K A_k(t) s_k(2\pi\phi_k(t)),
\end{equation}

\noindent where $A_k(t)$ and $\phi'_k(t)$ carry the same meaning as before but the cosine waves are replaced by more general oscillatory signals  $s_k(t)$, which are denoted as \emph{wave-shape functions} (WSFs). This model is known as the adaptive non-harmonic (ANH) model, in contrast with the AH model \eqref{eq:AH}. Each WSF $s_k$ is a $2\pi$-periodic, unit $L^2$ norm signal and can be described by its Fourier series expansion $s_k(t) = \sum_{\ell\in\mathbb{Z}} \hat{s}_k(\ell)e^{i\ell t}$, where $\hat{s}_k$ are the coefficients of the Fourier series corresponding to $s$. Formally, the WSFs belong to the analytic shape function class:
\begin{definition}[Analytic shape function class $\mathcal{S}^{\delta,D,\theta}$]
	\label{def:analytic_class}
    \doublespacing
	Given $\delta\geq 0$, $D\in \mathbb{N}$, and $\theta\geq 0$, the class $\mathcal{S}^{\delta,D,\theta}$ is defined as the $1$-periodic functions $s(t)$ with zero-mean ($\hat{s}(0)=0$), unit $L^2$-norm and satisfying:
	\begin{itemize}
		\item[S1.]
		$ \forall \ell \in \mathbb{Z}$,\ \text{with}\ $|\ell|\neq 1$, $|\hat{s}(\ell)| \leq \delta |\hat{s}(1)|,$
		\item[S2.] $\sum_{|\ell|>D} |\ell \hat{s}(\ell)|\leq \theta$.
	\end{itemize}
\end{definition}

Condition $\text{S1}$ indicates that the fundamental component cannot be zero. Moreover, if $\delta<1$, the first harmonic component of $s(t)$ is said to be \emph{dominant}. Condition $\text{S2}$ says that the coefficients decay fast enough as $\ell$ increases so that the wave-shape can be accurately approximated by a function with limited bandwidth. 
\subsection{Time-Varying Wave-Shape Model}
The ANH model has been applied to a variety of real-world scenarios, including the analysis of various physiological signals~\cite{lin2016sleep,su2017extract,li2019non}. Nevertheless, this model assumes that the WSF remains fixed throughout time. In many signals, including several biomedical signals, the oscillatory patterns are time-varying. Consequently, a more general oscillatory model is required. To achieve this, first consider the \emph{monocomponent} ANH model (i.e. $K=1$). If $s$ belongs to class $\mathcal{S}^{\delta,D,\theta}$, then we can replace it with its Fourier series expansion

\begin{equation}
\label{eq:ANH1.5}
    x_{\textsc{fix}}(t) = A(t) s(2\pi \phi(t)) = A(t)\sum_{\ell=1}^\infty a_\ell cos(2\pi\ell\phi(t) + \varphi_\ell).
\end{equation}

Here, the morphology of wave-shape function $s$ is characterized by the Fourier coefficients $a_\ell$ and the shifts $\varphi_\ell$. In~\cite{lin2018wave}, a generalization of \eqref{eq:ANH1.5}, which admits a time-varying WSF, is presented:

\begin{equation}
	\label{eq:ANH2}
	x_{\textsc{var}}(t) =  \sum_{\ell=1}^\infty B_\ell(t) \cos\left(2\pi \phi_\ell(t)\right).
\end{equation}

This model considers a non-stationary signal with time-varying wave-shape function as the superposition of ``harmonic" components $B_\ell(t)\cos(2\pi\phi_\ell(t))$ that satisfy the following conditions:

\begin{itemize}
	\item[(C3)]
	$\phi_\ell \in C^2(\mathbb{R})$ with $|\phi'_\ell(t) - \ell\phi'_1(t)|\leq \epsilon_3\phi'_1(t)$, for all $t$ and for $\ell = 1,2,\dots, \infty$.
	\item[(C4)] $B_\ell \in C^1(\mathbb{R})\cap L^\infty (\mathbb{R})$, with $B_\ell(t) > 0$ and $B_\ell(t)\leq c(\ell)B_1(t)$, with $\{c(\ell)\}_{\ell=1}^\infty$
	being a non-negative $\ell^1$ sequence, for all $t$ and for $\ell = 1,2,\dots, \infty$ .
	\item[(C5)] $|B'_\ell(t)|\leq\epsilon c(\ell)\phi'_1(t)$ and $|\phi''_\ell(t)|\leq\epsilon\ell\phi'_1(t)$ for all $t$ and $\text{sup}_{\ell;B_\ell\neq 0} \|\phi''_\ell\|_{\infty} = M$ for some $M\geq 0$. Moreover, there exist $N\in \mathbb{N}$ so that $\sum_{\ell=N+1}^\infty B_\ell(t)\leq \epsilon \sqrt{\sum_{\ell=1}^\infty B_\ell^2(t)}$ and $\sum_{\ell=N+1}^\infty \ell B_\ell(t)\leq C\sqrt{\sum_{\ell=1}^\infty B_\ell^2(t)}$. 
\end{itemize}

Condition (C3) indicates that the IF of each component is close to, but not exactly, an integer multiple of the fundamental frequency. Condition (C4) accounts for wave-shape variability given by the variation of the amplitude for each harmonic when the ratio $B_\ell(t)/B_1(t)$ is not constant. Condition (C5) establishes that the wave-shape varies \emph{slowly}, that is, there are no abrupt changes in the oscillatory pattern of the signal.  Note that we can recover the Fourier expansion of $s$ shown in \eqref{eq:ANH1.5} when $\phi_\ell(t) = \ell\phi(t) + \varphi_\ell/(2\pi)$ and $B_\ell(t) = A(t)a_\ell$ for $\ell \geq 1$.

In order to apply model \eqref{eq:ANH2}, we need to be able to characterize the time-varying nature of $\phi_\ell(t)$ and $B_\ell(t)/B_1(t)$. In~\cite{colominas2021decomposing}, an approach to solving the time-varying wave-shape model was proposed using the following simplified version of Eq. \eqref{eq:ANH2}

\begin{equation}
	\label{eq:SAMD}
	x_{\textsc{SAMD}}(t) =  A(t)\sum_{\ell=1}^\infty a_\ell\cos\left(2\pi \phi_\ell(t)\right),
\end{equation}

\noindent while satisfying conditions (C1), (C2), (C3) and (C5). We can see that model \eqref{eq:SAMD} is equivalent to \eqref{eq:ANH2} when $B_\ell(t) = A(t)a_\ell$, that is, when the amplitude modulation of the $\ell$-th harmonic component is a multiple of the instantaneous amplitude $A(t)$.  Without lose of generality, we can assume $B_1(t) = A(t)$ and consequently $a_1 = 1$. On the other hand, the relationship between the instantaneous phase of the harmonics and the fundamental component is modeled as a polynomial: $\phi_\ell(t) = \sum_{i}^I e_{i,\ell}\phi_1(t)^i$, with the $e_{i,\ell}$ being close to integers and considering $\varphi_\ell = 0$ for all $\ell$.

Generally, first-order polynomials are sufficient to give a good approximation of time-varying wave-shape so the equation reduces to $\phi_\ell = e_\ell\phi_1(t)$. In short, the WSF variability is characterized by the coefficients $e_{\ell}$, which are close to $\ell$ due to condition (C3). However, to fully capture the temporal variations in the wave-shape, we need to somehow incorporate condition (C4) into the model. 

In this work, we propose a new approach to wave-shape estimation where the variability of the waveform is considered both by non-integer phases, as well as the time variation in the amplitude ratio of the harmonics ($B_\ell(t)/B_1(t)$). Our proposal starts from \eqref{eq:ANH2} by first expanding the fundamental component in the summation
\begin{align}
	\label{eq:tvWSE}
	x(t) = B_1(t)\cos\left(2\pi\phi_1(t)\right) + \sum_{\ell=2}^\infty B_\ell(t)\cos\left(2\pi\phi_\ell(t)\right),
\end{align}

\noindent where $B_\ell(t)$ is the IA of the $\ell$-th component of the signal. We see that the first term $B_1(t)\cos(2\pi\phi_1(t))$ is the fundamental component of $x(t)$ and, when compared to $\eqref{eq:SAMD}$, we identify $B_1(t)=A(t)$. This fundamental component encodes the global variation of the amplitude and frequency of $x(t)$. In general, the global amplitude modulation affects all harmonics of $x(t)$, so we divide the signal by $B_1(t)$ to get the following 
\begin{equation}
\label{eq:tvWSEN}
\Tilde{x}(t) = \frac{x(t)}{B_1(t)} =\cos\left(2\pi\phi_1(t)\right) + \sum_{\ell=2}^{\infty}\underbrace{\frac{B_\ell(t)}{B_1(t)}}_{Q_\ell(t)} \cos\left(2\pi\phi_\ell(t)\right),
\end{equation}
\noindent which results in an \emph{amplitude-demodulated} time-varying wave-shape signal. Finally, the time-varying wave-shape is encoded in the time evolution of the harmonic amplitude functions $Q_\ell(t) = B_\ell(t)/B_1(t)$ and harmonic phase functions $\phi_\ell(t)$, i.e. the wave-shape function does not change if the $Q_\ell(t)$ functions are constant and if $\phi_\ell(t) = \ell \phi_1(t)$, with $\ell = 2,3,\dots$. In general, the phase of each component may contain a constant phase shift $\varphi_\ell$ that differs for each harmonic. In order to account for this, we rewrite the term $\cos(2\pi\phi_\ell(t))$ into the sum of a sine wave and a cosine wave with a new phase function $\Phi_\ell(t)= \phi_\ell(t) - \varphi_\ell$. Additionally, we set $\Phi_\ell(t) = e_\ell\phi_1(t)$ by following the proposal in \cite{colominas2021decomposing} using first order polynomials for the approximation of the phase functions, resulting in
\begin{align}
	\Tilde{x}(t) = \cos\left(2\pi\phi_1(t)\right) + \sum_{\ell=2}^\infty & Q_\ell(t) \left(a_\ell\cos\left(2\pi e_\ell\phi_1(t)\right) + b_\ell\sin\left(2\pi e_\ell\phi_1(t)\right) \right).
\end{align}
We note here that we assume $\varphi_1=0$ and, consequently, $b_1=0$. This introduces two new coefficients, $a_\ell$ and $b_\ell$, to be estimated for each harmonic. A further refinement of the model can be made by doing
\begin{flalign}
	\label{eq:tvWSE2}
	\Tilde{x}(t) & = \cos(2\pi\phi_1(t)) + \sum_{\ell=2}^\infty Q_\ell(t)a_\ell  \left(\cos\left(2\pi e_\ell\phi_1(t)\right) + \frac{b_\ell}{a_\ell}\sin\left(2\pi e_\ell\phi_1(t)\right)\right).
\end{flalign}
\noindent Finally, setting $\alpha_\ell(t) = Q_\ell(t)a_\ell$ and $c_\ell = a_\ell/b_\ell$ we get
\begin{flalign}
	\label{eq:tvWSE3}
	\tilde{x}(t) & = \cos(2\pi\phi_1(t)) + \sum_{\ell=2}^\infty \alpha_\ell(t) \left(\cos\left(2\pi e_\ell\phi_1(t)\right) + c_\ell\sin\left(2\pi e_\ell\phi_1(t)\right)\right).
\end{flalign} 
This way, the constant phase shift $\varphi_\ell$ is encoded by the coefficients $c_\ell$. We will refer indistinctly to both $Q_\ell(t)$ and $\alpha_\ell(t)$ as the harmonic amplitude function (HAF) given that they differ only by a scaling constant. 

\subsubsection{Adaptive Time-Varying Wave-Shape Extraction Algorithm}
\label{sec:tvWSE}
The proposed algorithm for wave-shape estimation considers the following least-squares optimization problem: 

\begin{equation}
	\label{eq:nonlinreg_cont}
	\underset{\scriptsize \begin{matrix}\alpha_{\ell}, c_\ell\\ e_\ell \end{matrix}}{\min} \ \left\|\tilde{x}(t)- \sum_{\ell=1}^r \alpha_\ell(t)\Theta_\ell(t)\right\|^2,
\end{equation}

\noindent where $\alpha_\ell(t)$ is the HAF for the $\ell$-th harmonic component of the time-varying wave-shape and $\Theta_\ell(t) = \cos\left(2\pi e_\ell\phi_1(t)\right) + c_\ell\sin\left(2\pi e_\ell\phi_1(t)\right)$. We set $\alpha_1(t) = 1$, $c_1 = 0$, and $e_1=1$. This formulation leads to a non-linear regression (NLR) on the set of parameters $\{\alpha_\ell(t),c_\ell,e_\ell\}_{\ell=2}^r$. Details on the resolution of this NLR problem are postponed until the numerical implementation section. Note here that we only sum the first $r$ components of the (phase-modulated) time-varying wave-shape. This is based on condition $S1$ for the analytic shape function class (see Def. \ref{def:analytic_class}), where it is reasonable to assume that the WSF can be accurately estimated as a band-limited function. The parameter $r$ needs to be chosen before solving the nonlinear regression problem. To do this, we use model selection criteria based on trigonometric regression~\cite{eubank1990curve,wang1993aic}, originally developed for stationary trigonometric time series. Nevertheless, results in~\cite{ruiz2022wave} show that they can be applied in the case of non-stationary signals to obtain an accurate estimate of the number of harmonic components of the WSF. 

A key aspect of this model is how to characterize the variations of the harmonic amplitude functions $\alpha_\ell(t) = Q_\ell(t)a_\ell$. A complete description of each harmonic amplitude for each time instant $t$ will lead to an overly redundant representation of the information contained in $x(t)$ and greatly increase the computational load. Therefore, a compact representation of the variability of the harmonic amplitudes is required. This consideration motivates the following definition of the harmonic amplitude function. 

\begin{definition}[Harmonic Amplitude Function]
	\label{def:HAF}
	Given $\ell\geq 2$, a harmonic amplitude function $\alpha_\ell(t) = Q_\ell(t)a_\ell$ as given in Eq. \eqref{eq:tvWSE3} is a $\mathbb{C}^1$ function defined on a closed interval: $\mathbb{I}\longrightarrow\mathbb{R}$, with finite $L^2$ norm and 
	\begin{align*}
		\|P_\ell(t) - \alpha_\ell(t) \|<\epsilon_4,\ \text{for}\ t\in\mathbb{I},
	\end{align*}
	
	\noindent for a small $\epsilon_4>0$, where $P_\ell(t)$ is an interpolating function using $I_\ell$ nodes noted as $t_i$, for $i=1,\dots,I_\ell$.
\end{definition}

Definition $\ref{def:HAF}$ indicates that the HAFs can be accurately estimated from a set of $I_\ell$ nodes, located at times $t_{\ell,i}$ and with amplitudes $\alpha_{\ell,i} = \alpha_\ell(t_{\ell,i})$, using an interpolating function $P_\ell(t)$, which is normally defined as a piecewise function depending on the nodes amplitude and location 

\begin{equation}
	P_\ell(t) = \sum_{i=1}^{I_\ell-1} P_{\ell,i}(t),
\end{equation}

\noindent where each $P_{\ell,i}(t)$ is a compactly-supported function defined in the subinterval $[t_{\ell,i}, t_{\ell,i+1}]$ and $P_{\ell,i}(t_i) = \alpha_{\ell,i}$, for all nodes. Additional conditions on the continuity (first derivative) and concavity (second derivative) of the interpolation curve may be desired and will affect the choice of interpolation functions. In this work, we used shape-preserving cubic Hermite interpolators~\cite{fritsch1980monotone} for HAFs interpolation. We will discuss the details of this choice in the next section.

\section{Numerical Implementation}
\label{sec:numerical}
The model estimation algorithm herein proposed requires the discretization of the signal. To do this, we set a sampling frequency $f_s$ in Hz (samples/seconds) and sample $x(t)$ at a regular interval given by the sampling rate $\Delta t=1/f_s$ to obtain the discrete signal $\mathbf{x}(n) = x(n\Delta t)$, with $1\leq n \leq N$. This discrete signal has a finite duration of $T = N/f_s$ seconds. Given \eqref{eq:tvWSE}, the amplitude modulation, phase modulation, and harmonic amplitude functions will also be discrete: $\boldsymbol{\phi}_1(n) = \phi_1(n\Delta t)$, $\mathbf{B}_1(n)=B_1(n\Delta t)$ and $\boldsymbol{\alpha}_\ell(n)=\alpha_\ell(n\Delta t)$.

\subsection{Discrete Encoding of Harmonic Amplitude Functions}
The harmonic amplitude functions encode relevant information about the time-varying pattern of the wave-shape. As mentioned in the previous section, HAFs are smooth functions of time that govern the evolution of the wave-shape through time. Given this, we carry out a compressed encoding of the HAF with minimal loss of information through free-node interpolation using cubic interpolators. For each harmonic, we define a set of node values $\{\alpha_{\ell,i}\}$ and their corresponding time locations $\{t_{\ell,i}\}$, so that $\alpha_{\ell,i}=\alphab_\ell(t_{\ell,i})$ for $i=1,\dots,I_\ell$. Generally, the number of nodes $I_\ell$ can differ for each harmonic component. The nodes at the start and end of the signal are fixed $t_{\ell,1}=0$ and $t_{\ell,I_\ell}=T$ for all $\ell\geq 2$. The inner nodes are free to vary in their location given the restrictions $0<t_{\ell,i}<T$ and $t_{\ell,i-1}<t_{\ell,i}<t_{\ell,i+1}$ for all $i=2,\dots,I_\ell-1$.

Given the set of nodes amplitudes $\{\alpha_{\ell,i}\}$ and locations $\{t_{\ell,i}\}$, the HAFs are approximated by piecewise interpolation $\boldsymbol{\alphah}_\ell(n)= \sum_{i=1}^{I_\ell-1} \mathbf{P}_{\ell,i}(n)$, where $\mathbf{P}_{\ell,i}(n)$ is the shape-preserving cubic Hermite interpolator defined for the subinterval $I_{\ell,i} = [t_{\ell,i},t_{\ell,i+1}]$ and $\mathbf{P}_{\ell,i}(t_{\ell,i})=\alpha_{\ell,i}$. This type of shape-preserving interpolator is monotone in the interval $[t_{\ell,1},t_{\ell,I_\ell}]$ if the node amplitude sequence $\{\alphab_{\ell,i}\}$ is also monotone. Additionally, if $\alphab_\ell(n)$ has a local extremum at $t_{\ell,i}$ then $\mathbf{P}_{\ell,i}(n)$ also does. Shape-preserving cubic interpolators have continuous first derivatives in the nodes, but the continuity of the second derivative is not guaranteed. 

\textbf{Remark:} \emph{On monotonicity and concavity of harmonic amplitude functions}: Given the Definition $\ref{def:HAF}$, we require an interpolation scheme that recovers the time evolution of the HAF as faithfully as possible, given the slow-varying and smooth conditions. Moreover, we should expect that for a monotone and convex curve, the resulting interpolation curve is also monotone and convex. To achieve this, shape-preserving cubic interpolators are useful given that they preserve the monotonicity and concavity of the interpolated curve. Other commonly used interpolators, like cubic spline, result in curves with unnatural wiggles and bumps in between the nodes which affect the accuracy of the estimated curve. 

\subsection{Optimal Node Number Estimation}
The number of interpolation nodes $I_{\ell}$ is an important parameter that can vary from one harmonic to another and greatly affects the accuracy of the estimated curve. Therefore, automatic estimation of $I_{\ell}$ needs to be performed. To do this, we first note that for a signal modeled as in \eqref{eq:tvWSE3}, a rough estimate of the complexified version of the $\ell$-th harmonic $\hat{y}_\ell(t)$ can be obtained by vertical reconstruction of the short-time Fourier transform (STFT):

\begin{equation}\label{eq:reconstruction}
	\hat{y}_\ell(t) = \frac{2}{g(0)} \int_{|f-c_\ell(t)|<\Delta_\ell}F_x^g(t,f) df,
\end{equation}

\noindent where $\Delta_\ell$ is the reconstruction bandwidth for the $\ell$-th harmonic, and $c_\ell(t)\approx \phi'_\ell(t)$ is the estimation of the instantaneous frequency curve corresponding to the $\ell$-th ridge of $x(t)$. In essence, $|\hat{y}_\ell(t)|$ is a rough estimate of the modulus of the HAF of the $\ell$-th harmonic. Then, the optimal number of nodes for the HAF encoding can be estimated by considering the modulus $|\hat{y}_\ell(t)|$ as a band-limited signal with bandwidth $BW_\ell$. To obtain this bandwidth, the harmonic amplitude energy spectrum (ES) is computed by squaring the modulus of $\hat{Y}_\ell(f) = \mathcal{F}\{\hat{y}_\ell(t)\}$, where $\mathcal{F}$ is the Fourier transform operator. 

The spectrum bandwidth $\operatorname{BW}_\ell$ can be estimated as the $f$ coordinate where the normalized cumulative spectral energy

\begin{equation}
    CSE(f) = \frac{\int_0^f |\hat{Y}_\ell(\xi)|^2 d\xi}{\int_0^\infty |\hat{Y}_\ell(\xi)|^2 d\xi)},
\end{equation}

\noindent reaches $90\%$ of the energy of the harmonic amplitude estimate $|\hat{y}_\ell(t)|$. Using the Nyquist theorem, we then set $I_\ell = 2\operatorname{BW}_\ell +1$. 

\subsection{Model Coefficients Estimation}
Given the discrete signal $\mathbf{x}(n)$, we want to find the optimal time-varying wave-shape coefficient vector 
\begin{equation}
\boldsymbol{\gamma} = \left[\{t_{\ell,i}\}_{i=2}^{I_\ell-1},\{\alpha_{\ell,i}\}_{i=1}^{I_\ell},c_\ell,e_\ell\right]_{\ell=2}^r \in \mathbb{R}^H,
\end{equation}

\noindent with $H = 2\sum_{\ell=2}^r I_\ell$. This optimal coefficient vector can be estimated by solving the least-squares non-linear regression problem \eqref{eq:nonlinreg_cont} in discrete time:

\vspace{-0.15in}
\begin{equation}
	\label{eq:nonlinreg}
	\hat{\gammab} =     \underset{\gammab\in \mathbb{R}^H}{\min}\ \left\|\tilde{\mathbf{x}}(n)- \left[\sum_{\ell=1}^r \alphab_\ell(n)\odot \boldsymbol{\Theta}_\ell(n)\right]\right\|^2,
\end{equation}
where $\alphab_\ell(n)$ is the result of interpolating the nodes amplitudes $\{\alpha_{\ell,i}\}$ at node locations $\{t_{\ell,i}\}$ using shape-preserving cubic interpolators and $\boldsymbol{\Theta}_\ell(n) = \cos\left(2\pi e_\ell\phib_1(n)\right) + c_\ell\sin\left(2\pi e_\ell\phib_1(n)\right)$, for $n = 1,2,\dots,N$. As mentioned before, this regression problem is nonlinear on the model coefficients given Eq. \eqref{eq:nonlinreg}. The optimal solution is obtained by using an iterative non-linear curve fitting procedure based on the Levenberg-Marquardt algorithm~\cite{marquardt1963algorithm}.

Given the relationship $\alphab_\ell(n) = \mathbf{Q}_\ell(n)a_\ell$, we deduce that $\mathbf{Q}_\ell(n)$ encodes the amplitude variability of the demodulated $\ell$-th harmonic. The HAFs $\alphab_\ell(n)$ encode information about the wave-shape variability and the relative harmonic amplitude.

The combination of the adaptive selection of the number of harmonics of the WSF, the automatic estimation of the number of nodes for the HAFs, and the non-linear regression leads to a fully adaptive time-varying wave-shape extraction algorithm. We summarize the steps of the fully adaptive procedure in the upper flowchart diagram of Fig. \ref{fig:flowchart}.
\subsection{Algorithm Initialization}
The initial conditions of the curve-fitting algorithm have to be consciously chosen to ensure the convergence of the algorithm to a reasonable solution. We propose a ``warm start'' initialization~\cite{yildirim2002warm} based on the solution of the linear regression (LR) approach for the ANH model with fixed wave-shape given in Eq. \eqref{eq:ANH}. This is based on the following LSE problem~\cite{wu2016modeling}:
\begin{align}
	\hat{\gammab}_{LR} = \underset{\gammab\in\mathbb{R}^{2r}}{\text{arg min}} \left\| \tilde{\mathbf{x}}(n) - \tilde{\mathbf{C}}_r \boldsymbol{\gamma} \right\|,
\end{align}

\noindent with $\tilde{\mathbf{C}} = [\tilde{\mathbf{c}}_1(n),\dots,\tilde{\mathbf{c}}_r(n),\tilde{\mathbf{d}}_1(n),\dots,\tilde{\mathbf{d}}_r(n)]$, $\tilde{\mathbf{c}}_\ell(n) = \cos(2\pi\ell\phib_1(n))$ and $\tilde{\mathbf{d}}_\ell(n) = \sin(2\pi\ell\phib_1(n))$. By direct optimization, using the normal equations, we obtain the optimal LR coefficient vector:

\begin{equation}
    \hat{\gammab}_{LR} = (\tilde{\mathbf{C}}_r^T\tilde{\mathbf{C}}_r)^{-1}\tilde{\mathbf{C}}_r^T \tilde{\mathbf{x}} = \left[\alphah_{LR,1},\dots,\alphah_{LR,r},\betah_{LR,1},\dots,\betah_{LR,r}\right].
\end{equation}

\begin{figure}
    \centering
    \includegraphics[width=\columnwidth]{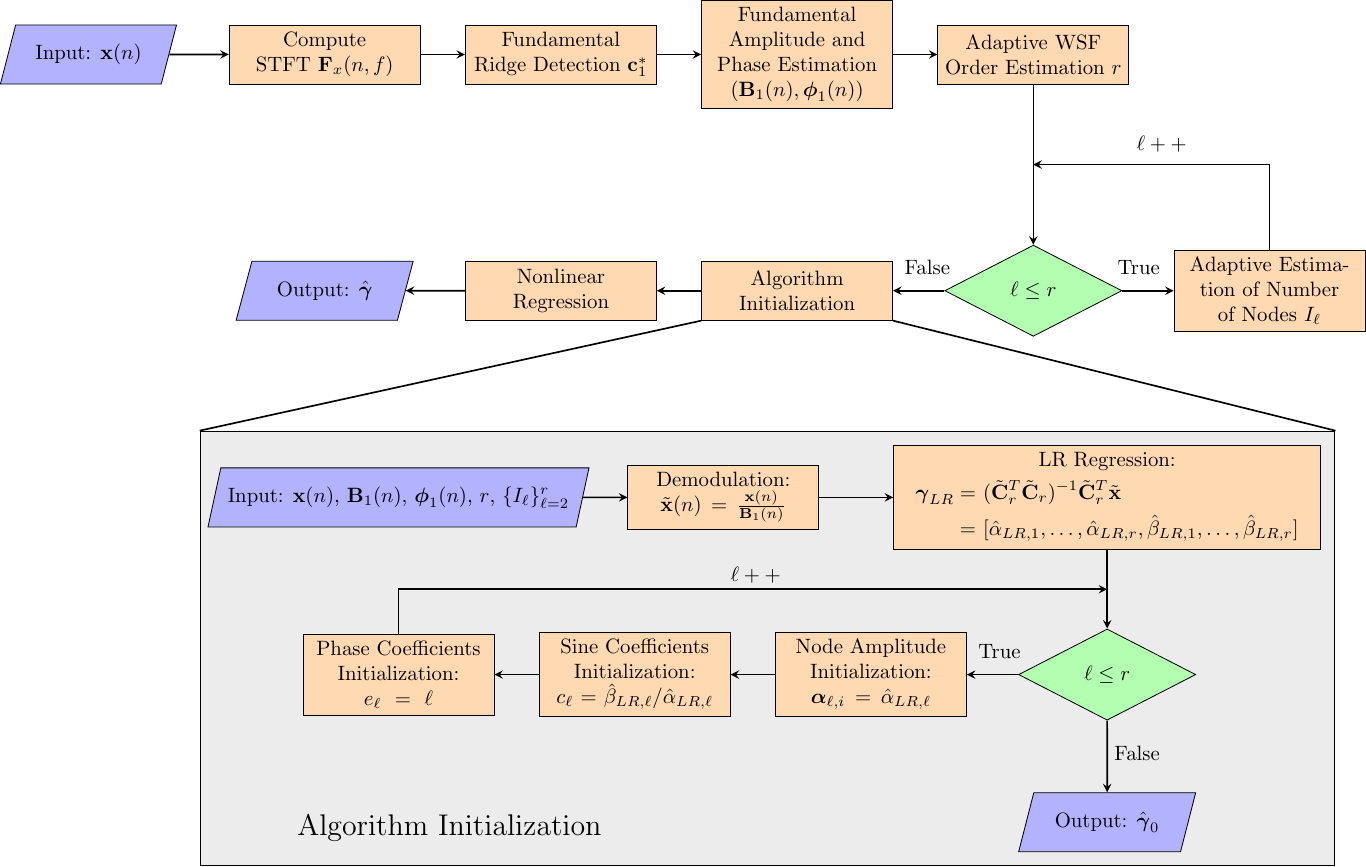}
    \caption{Flowchart for the adaptive time-varying wave-shape extraction algorithm. The flowchart detailing the steps of the initialization of the non-linear regression step is shown in the gray box.}
    \label{fig:flowchart}
\end{figure}

Then, initial node amplitudes $\{\alpha_{\ell,i}\}$ are set equal to the cosine linear regression coefficients $\hat{\alpha}_{LR,\ell}$ for all $I_\ell$ nodes in each harmonic. Initial node locations $\{t_{\ell,i}\}$ are set as an equidistant grid of $I_\ell$ points on the interval $[0,T]$. Coefficients $c_\ell$ are initialized as $\hat{\beta}_{LR,\ell}/\hat{\alpha}_{LR,\ell}$. Finally, harmonic phase coefficients are set to $e_\ell = \ell$ for $\ell = 2,\dots,r$. These coefficients are concatenated to form the initial parameter vector $\gammab_0$. The steps of the initialization procedure are shown in the flow chart at the bottom of Fig. \ref{fig:flowchart}.

The fully adaptive time-varying wave-shape extraction algorithm was implemented using MATLAB$^\copyright$ version R2021b. The code for the proposed method and all experiments detailed in this document are available at: \url{https://github.com/joaquinr-uner/tvWSE}.

\section{Experimental Results on Synthetic Signals}
\label{sec:synth}
In this section, we show the results of several experiments on simulated data for various signal processing tasks. For these experiments, without loss of generality, synthetic signals are limited to the $[0,1]$ interval and their mean is set to zero. When we apply our time-varying wave-shape extraction algorithm, particular care needs to be taken near the borders of the signal. When solving Eq. \eqref{eq:nonlinreg}, border effects manifest in two ways. Firstly, the analysis window used to compute the STFT affects the TF resolution at the borders of the spectrogram. These effects lower the accuracy of the AM and FM estimates at the start and end of the signal. Secondly, the interpolation of the HAFs nodes using shape-preserving cubic interpolation cannot specify the value of the first derivative of the interpolating curve at the extreme nodes $t_{\ell,1}$ and $t_{\ell,I_\ell}$, leading to inaccurate estimation of the node amplitudes at these instants. 

In order to overcome these effects, the signal is forward extended by fitting a seasonal ARIMA model~\cite{box2015time} to the last $3$ cycles of the signal and then using the fitted model to forecast the next $N_p = 0.1N$ samples. Preliminary experiments show that this value for $N_p$ gives a good trade-off between computation time and boundary effect attenuation. We then extend the signal in the backward direction by flipping it: $x^*(t)=x(-t)$, and applying the same procedure. Note that this extension also adds two extra fixed nodes at the start and end for each harmonic starting from $\ell = 2$, increasing the number of coefficients to be estimated for each signal to $2\sum_{\ell=2}^r I_{\ell} + 2(r - 1)$. The ARIMA forecasting procedure is chosen given that it is a straightforward method for series forecasting that is able to preserve the oscillatory pattern at the start and end of the signal.

For all experiments in this section, synthetic signals were discretized using a sampling frequency $f_s = 2000$ Hz. The amplitude and phase modulation estimates, $\mathbf{\hat{B}}_1(n)$ and $\hat{\phib}_1(n)$ respectively, were obtained from the reconstruction around the fundamental ridge of the STFT, according to \eqref{eq:reconstruction}. Note that we assumed that the fundamental component corresponds to the most energetic ridge in the spectrogram, which may not always be the case. Some tools, like the de-shape STFT~\cite{lin2018wave}, can be used to overcome this issue. Nevertheless, the dominance of the fundamental ridge was assumed here to simplify the analysis. To compute the STFT, we used a Gaussian window $\mathbf{g}(n) = e^{-\sigma n^2}$ with window variance $\sigma = 10^{-4}$. The fundamental ridge, noted as $\mathbf{c}(n)$, was obtained using the ridge extraction algorithm proposed in \cite{meignen2012new}. To obtain a smooth curve $\mathbf{c}(n)$, the aforementioned algorithm uses a greedy procedure where the maximum of the spectrogram is computed for time index $n$ and the search for the instant $n+1$ is restricted to the frequency band $[\mathbf{c}(n)-I_f,\mathbf{c}(n)+I_f]$, with $I_f$ being the maximum admissible frequency jump. The fundamental component was recovered by vertical reconstruction of the STFT limited to the frequency range $[\mathbf{c}(n)-\Delta,\mathbf{c}(n)+\Delta]$, where $\Delta$ was chosen to be close to the measure of the half-support of $\hat{g}(k) = \mathcal{F}(\mathbf{g}(n))$, the Fourier transform of the analysis window. Finally, the value of $r$ for the ANH model was estimated using the Wang selection criterion as described in \cite{ruiz2022wave}.

We compared our proposal to the LR and SAMD algorithms, which are based on Eq. \eqref{eq:ANH} and \eqref{eq:SAMD}, respectively. The former considers a fixed wave-shape while the latter allows for changes in the wave-shape. Implementation details for these methods can be found in~\cite{wu2016modeling} and~\cite{colominas2021decomposing}, respectively. Both LR and SAMD have been considered for the tasks of denoising and decomposition of signals with wave-shape functions. These methods were chosen because they serve as a well-established benchmark for the different tasks considered in this work. In particular, our proposal results in a more versatile version of the ANH model by considering all sources of wave-shape variability in its implementation. The three models were applied to synthetic signals for the tasks of time-varying wave-shape estimation, decomposition and denoising of both monocomponent and multicomponent signals. Another time-varying wave-shape estimation method, MMD, was not considered in these experiments due to its slow convergence rate, which is two orders of magnitude higher than the other methods.

\begin{figure}
	\centering
	\includegraphics[width=\columnwidth]{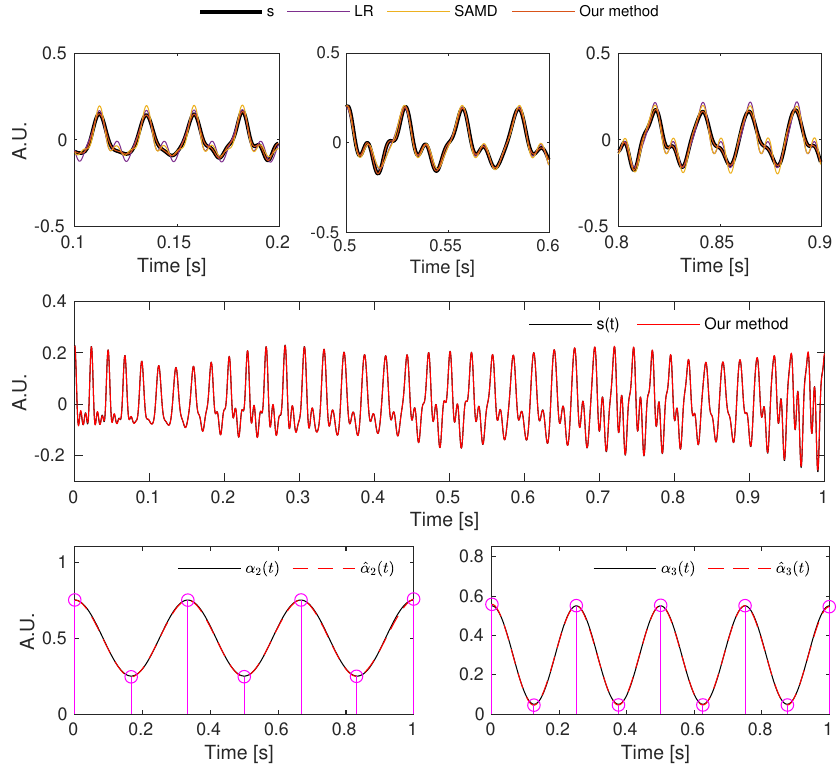}
	\caption{Top: $100$ millisecond-long segments at the start, middle and end of the signals highlighting the time-varying waveform with the results of LR, SAMD and our method superimposed. Middle: original signal (black) and the reconstruction using our method (red). Bottom: harmonic amplitude functions (full line), HAF estimation using our algorithm (dashed line) and node locations and amplitude (dots).}
	\label{fig:tvWSE1}
\end{figure}
\subsection{Reconstruction of Signals with Time-Varying Wave-Shape}
To show the capacity of our algorithm in recovering the information associated with the HAFs, we considered the following time-varying wave-shape signal
\begin{equation}
	x(t) = B_1(t)\left(\sum_{\ell=1}^3 \alpha_\ell(t)\cos\left(2\pi\phi_\ell(t)\right)\right),
\end{equation}

\noindent with $B_1(t) = 0.1\sqrt{t+1}$, $\phi_1(t) = 40t + \frac{5}{2\pi}\sin(2\pi t)$, $\alpha_1(t) = 1$, $\alpha_2(t) = 0.5 + 0.25\cos(2\pi 3t)$, and $\alpha_3(t) = 0.3 + 0.25\cos(2\pi 4t)$. The harmonic phases are $\phi_2(t) = 2.005\phi_1(t)$ and $\phi_3(t) = 2.995\phi_1(t)$.  The signal is shown in the middle panel of Fig. \ref{fig:tvWSE1} with our result superimposed in red. We can see that the wave-shape changes with time and the HAFs $\alpha_\ell(t)$ and phase functions $\phi_\ell(t)$ encode this variation. The close match between the original signal and our estimate shows how our algorithm is able to recover the information associated with the time-varying wave-shape of $x(t)$. Indeed, by looking at the top panels of Fig. \ref{fig:tvWSE1} we see that the output of LR and SAMD cannot fully capture the time evolution of the wave-shape. In the bottom row of Fig. \ref{fig:tvWSE1}, we show the theoretical HAFs $\alpha_2(t)$ and $\alpha_3(t)$, alongside the optimal nodes obtained with our adaptive algorithm and the interpolated curve. The optimal node amplitudes and locations are also shown alongside the interpolated curves plotted with dashed lines. We see that the time-varying pattern of the harmonic amplitudes is correctly recovered and border effects are negated due to the signal extension.

\subsection{Denoising Signals with Time-Varying Wave-Shape}
In this section, we evaluate the quality of our wave-shape extraction procedure in the presence of noise. To this end, we generated $200$ realizations of the noisy signal $x_n(t) = x(t) + n(t)$, where $n(t)$ is zero-mean Gaussian noise. The noise was characterized by the input signal-to-noise ratio: $\operatorname{SNR}_{in} = 20\log\left(\|x(t)\|/\|n(t)\|\right)$. 

\begin{figure}
	\centering
	\includegraphics[width=1\columnwidth]{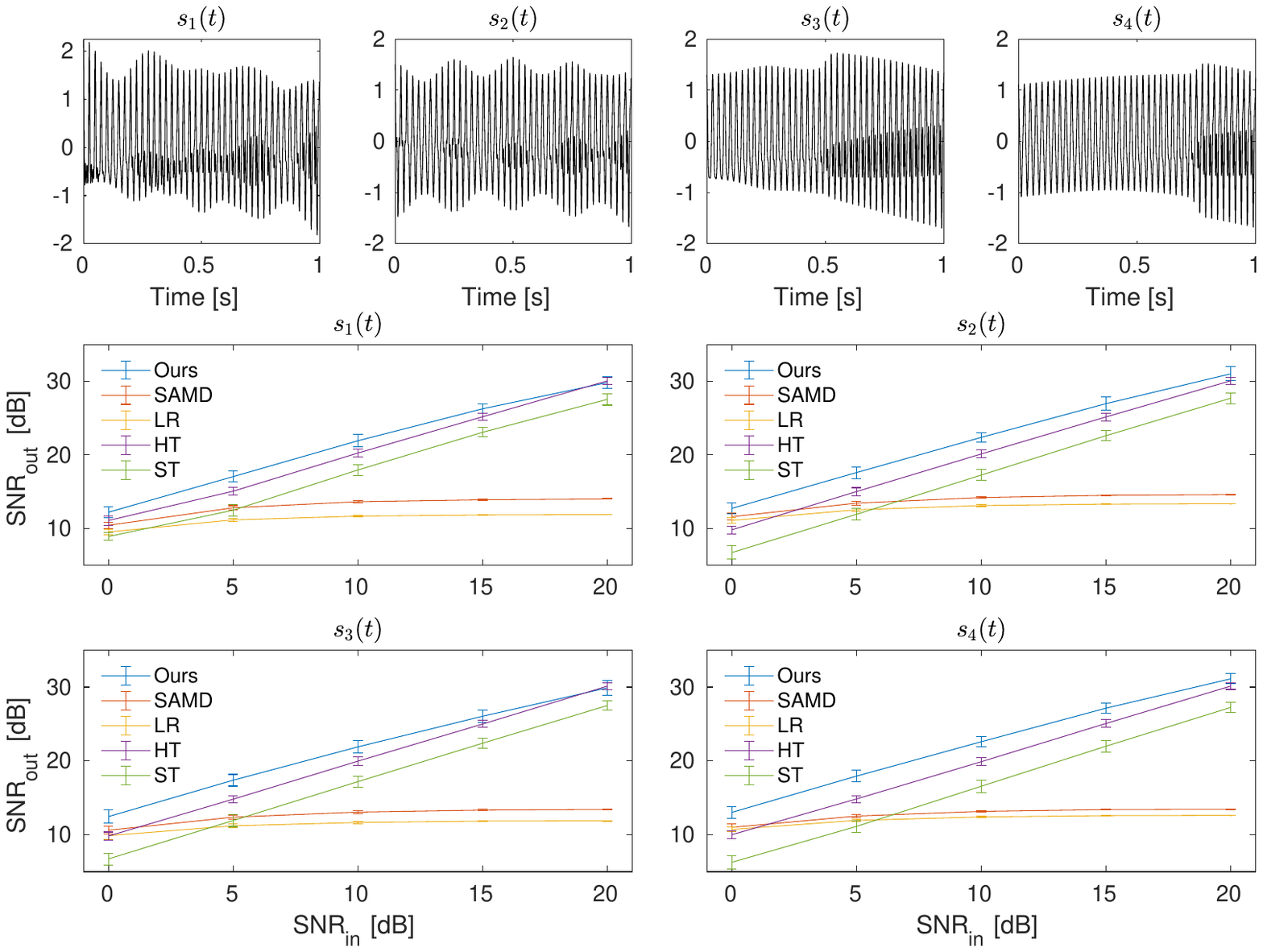}
	\caption{Top: Four different time-varying wave-shape functions. $s_1(t)$ - second and third HAFs are cosine modulated. $s_2(t)$ - second and third HAFs have linear and cosine modulation, respectively. $s_3(t)$ - second HAF has a hyperbolic tangent (tanh) modulation and the third HAF has a bump (Gaussian) modulation. $s_4(t)$ - second and third HAFs have linear and tanh modulation, respectively. Middle and bottom: Results of the denoising task for each denoising method for each type of time-varying WSF. References: LR - Linear regression - SAMD: Shape-adaptive mode decomposition - Ours: Adaptive time-varying wave-shape extraction - HT: Thresholded STFT with hard threshold. ST: Thresholded STFT with soft threshold.}	
    \label{fig:waveforms}
\end{figure}

For this experiment, 5 different input SNR levels were considered: $0$, $5$, $10$, $15$ and $20$ dB. In addition to the signal shown in Fig. \ref{fig:tvWSE1}, we studied a broader family of harmonic amplitude functions. The top row of Fig. \ref{fig:waveforms} shows a variety of signals with time-varying wave-shapes, given by different combinations of HAFs. The first two columns show signals with a smooth, slow-varying wave-shape given by sinusoidal and polynomic HAFs. The signals in the third and fourth columns have more abrupt changes in the wave-shape given by the faster rate of change of the hyperbolic tangent (tanh) and bump amplitude modulations. All signals considered in this section are of the form $x_i(t) = B_1(t)s_i(2\pi\phi_1(t))$ for $i=1,2,3,4$, using the same $B_1(t)$ and $\phi_1(t)$ from the previous section.

The performance of our algorithm was compared again to the LR and SAMD reconstructions. Additionally, we compared these adaptive non-harmonic model methods with the denoising procedure by thresholding of the STFT proposed in~\cite{pham2018novel}, using both hard and soft thresholding. Thresholding is a simple denoising operator that is commonly used for non-stationary signals. Moreover, given an adequate threshold, this technique is proven to reach close to optimal results with respect to minimizing the maximum risk of the signal estimator~\cite[Theorem 11.7]{mallat2008wavelet}. To implement this technique, we chose the threshold based on the noise variance estimate given in~\cite{donoho1994ideal}
\vspace{-0.05in}
\begin{equation}
	\hat{\sigma}^2 = \left(\frac{\text{median}\left(\left|\Re(\textbf{F}_\textbf{x}^\textbf{g}(n,k))\right|\right)}{0.6745\|\textbf{g}\|_2}\right)^2,
	\label{eq:thresh}
\end{equation}
where $\mathbf{F}_\mathbf{x}^\mathbf{g}$ is the STFT of signal $\mathbf{x}$ using a Gaussian window $\mathbf{g}(n)$. The optimal denoising threshold was set as $\eta = 3\sqrt{2}\hat{\sigma}\|\mathbf{g}\|_2$. The performance of each method was measured using the output SNR which is computed by $\operatorname{SNR}_{out} = 20\log\left(\| x_i(t)\|/\|\hat{x}_i(t)-x_i(t)\|\right)$, where $\hat{x}_i(t)$ is the reconstructed signal for the $i$-th time-varying wave-shape. The term $\|\hat{x}_i(t)-x_i(t)\|$ directly measures the estimation error. The output SNR was computed for all realizations for each time-varying wave-shape function using the five denoising methods.

\begin{figure}
	\centering
	\includegraphics[width=\columnwidth]{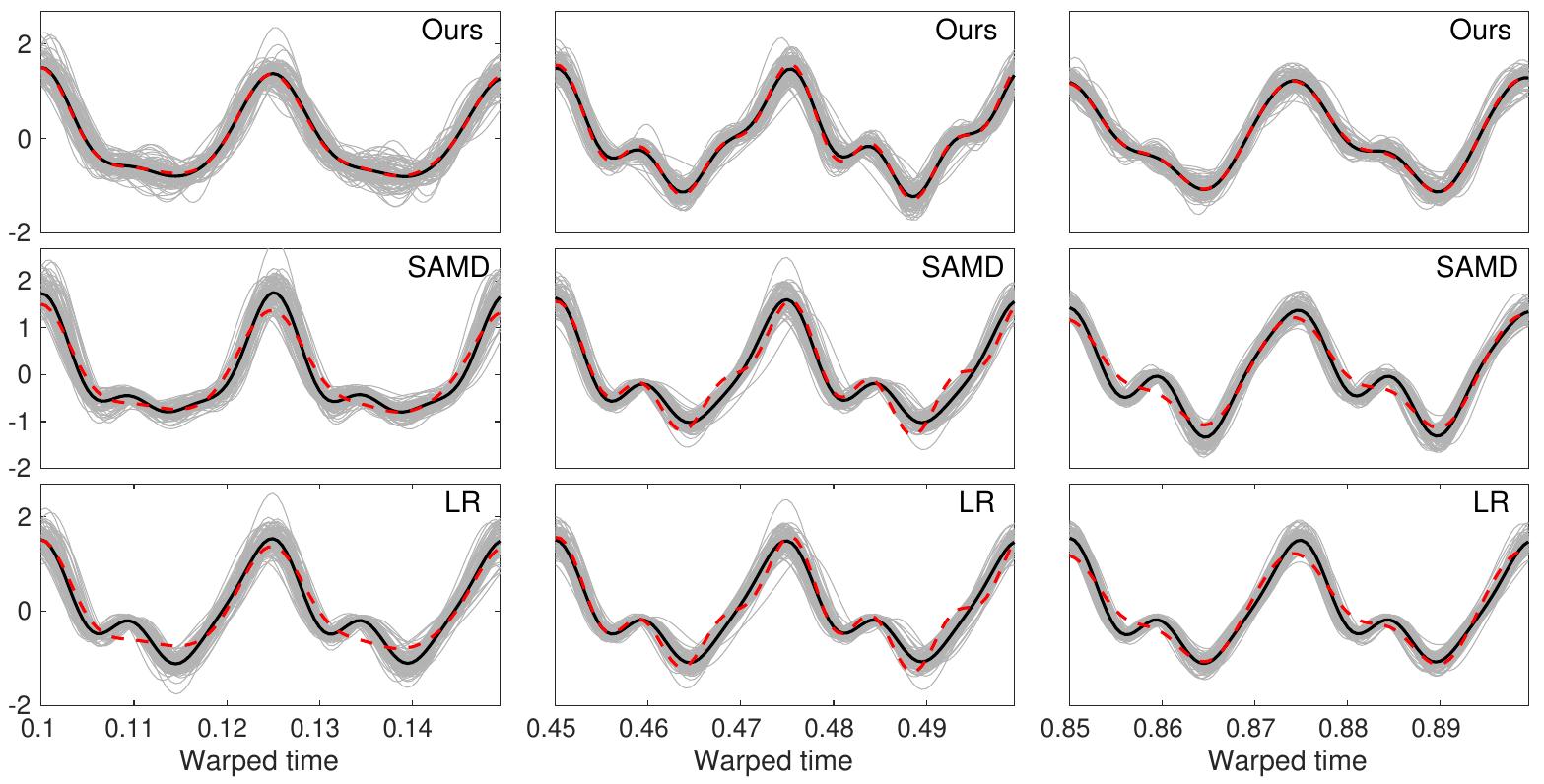}
	\caption{Extracted wave-shapes using our algorithm (top), SAMD (middle) and LR (bottom). In each row, $50$ millisecond-long windows at the start, middle and end of the $200$ realizations of the denoised and demodulated signal $\tilde{x}_{d,1}(t)$, with $\operatorname{SNR}_{in} = 0$  dB, are plotted in grey. The average wave-shape over all realizations is plotted in black and the original noiseless signal $s_1(t)$ is plotted using red dashed line.}
	\label{fig:denoisecomp}
\end{figure}

In the second and third rows of Fig. \ref{fig:waveforms} the $\operatorname{SNR}_{out}$ for each method, averaged over all realizations, is plotted as a function of $\operatorname{SNR}_{in}$ alongside the standard deviation interval. Firstly, we note that at low SNR levels ($0$ dB), the three ANH model reconstruction methods and the hard-thresholded STFT method have similar performance. The soft-thresholded STFT method has a considerably lower performance than the rest at this SNR level. The lower performance of the soft-threshold method is to be expected. As discussed in~\cite{mallat2008wavelet} (p. 556), the soft-threshold removes the TF coefficients with amplitude close to the threshold, whereas the hard-threshold leaves these coefficients unchanged. As the $\operatorname{SNR}_{in}$ increases, our method and the hard-thresholding method tend to obtain similar results, with the former slightly overcoming the latter. LR and SAMD show a quick saturation in performance as they cannot follow the time variations of wave-shape accurately. In the range from $0$ dB to $15$ dB our method outperforms all other methods. At $20$ dB, our method reaches an average $\operatorname{SNR}_{out}$ of $30$ dB, which is comparable to the results obtained by hard-thresholding.

We also compared the wave-shape estimation performance of the algorithms based on the ANH model. To do this, we demodulated the reconstructed signals $\hat{x}_i(t)$ in both amplitude and phase to obtain the demodulated signal $\hat{x}_{d,i}(t) = \hat{x}_i\left(\phi_1^{-1}(t)/(2\pi)\right)/B_1\left(\phi^{-1}(t)/(2\pi)\right)$. Given sufficiently accurate estimators of the fundamental amplitude and phase, the demodulated signal is close to the underlying time-varying wave-shape $\hat{x}_{d,i}(t) \approx s_i(t)$. Fig. \ref{fig:denoisecomp} shows the resulting time-varying wave-shape estimations for our method, LR and SAMD for $s_1(t)$ contaminated with $0$ dB noise. The mean curve for each method and the ground truth are plotted in thick black line and dashed red line, respectively. In summary, LR performs poorly as it considers a fixed waveform, which limits its capacity to accurately estimate the signal with a time-varying wave-shape. Both SAMD and our method capture the time variation of the waveforms and we see how the latter follows them more accurately by considering the time evolution of the harmonic amplitudes.

\begin{figure}
	\centering
	\includegraphics[width=1\columnwidth]{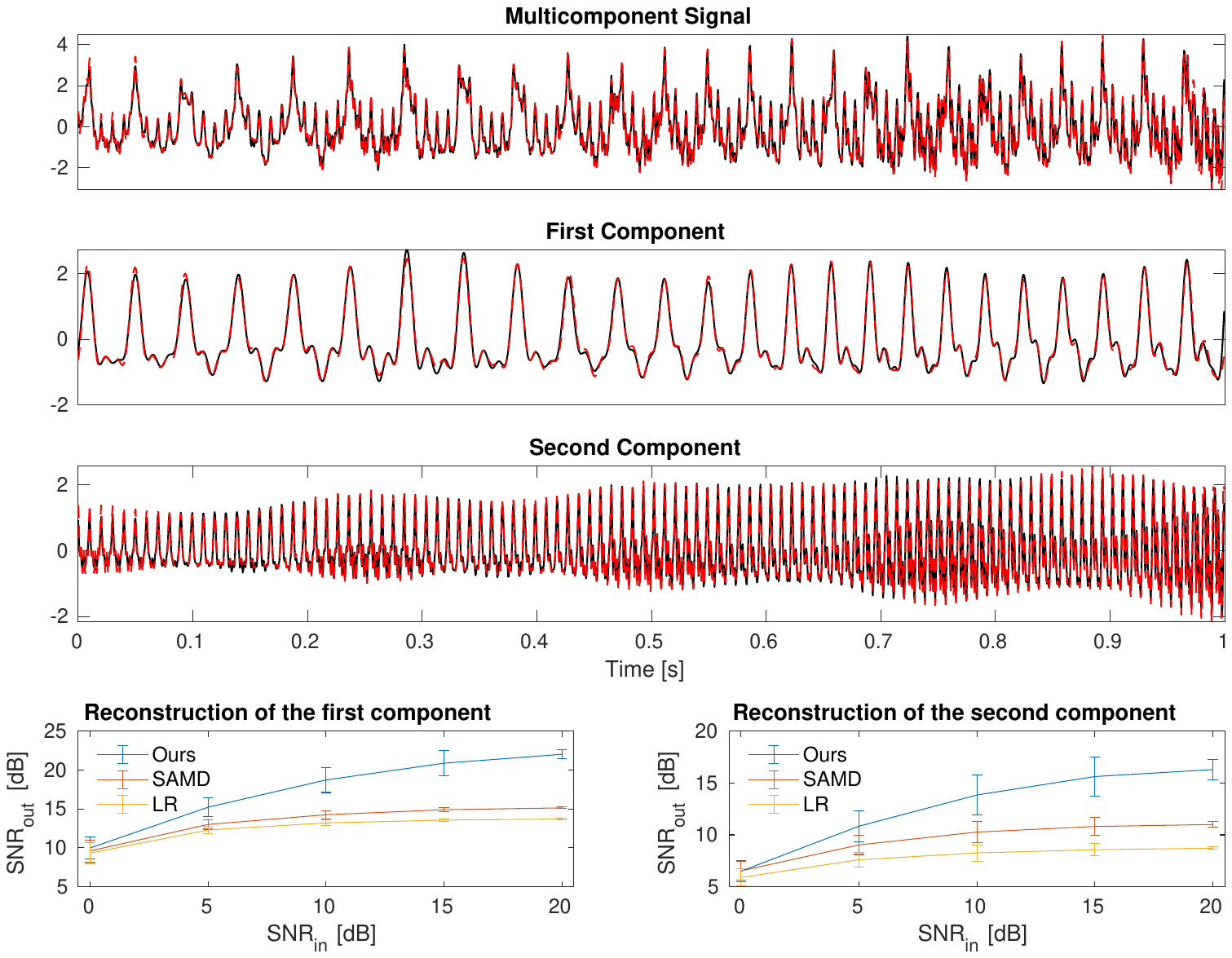}
	\caption{First row: Multicomponent signal $x(t)$ plotted in black with the reconstruction result for $\operatorname{SNR}_{in} = 10$ dB superimposed in dashed red line. Second row: Same for the first component signal $x_1(t)$. Third row: Same for second component signal $x_2(t)$. Fourth row: $\operatorname{SNR}_{out}$ as a function of $\operatorname{SNR}_{in}$ for all wave-shape decomposition methods for both component signals.}
	\label{fig:multicomp}
\end{figure}
\subsection{Decomposing Signals with Multiple Time-Varying Wave-Shapes}

We now consider the task of multicomponent signal decomposition when the component signals have time-varying WSFs. For this analysis, we denote the multicomponent signal as $x(t) = x_1(t) + x_2(t)$, where each component has a time-varying wave-shape. Indeed, $x_1(t)$ is defined as
\begin{equation}
	x_1(t) = B_{1,1}(t)\sum_{\ell=1}^3 \alpha_{1,\ell}(t)\cos\left(2\pi e_{1,\ell}\phi_{1,1}(t)\right).
\end{equation}

\noindent with $B_{1,1}(t) = \sqrt{0.01t} + 1.1$, $\phi_{1,1}(t) = 25t + \frac{5}{2\pi}\cos(2\pi t)$, $\alpha_{1,1}(t) = 1$, $\alpha_{1,2}(t) = 0.5 + 0.2\cos(2\pi 3 t)$, $\alpha_{1,3}(t) = \ 0.3 + 0.2e^{-(\frac{t-0.25}{0.1})^2}$,
$e_{1,2} = 2.005$ and $e_{1,3} = 3.003$. Similarly, the second component $x_2(t)$ is
\begin{equation}     
	x_2(t) = B_{2,1}(t)\sum_{\ell=1}^4 \alpha_{2,\ell}(t)\cos\left(2\pi e_{2,\ell}\phi_{2,1}(t)\right),
\end{equation}

\noindent with $B_{2,1}(t) = 2\log(t+1.1) + 0.5$, $\phi_{2,1}(t) = 100t + 7t^2$, $\alpha_{2,1}(t) = 1$, $\alpha_{2,2}(t) = \ 0.6 + 0.3t^2$, 
$\alpha_{2,3}(t) = 0.4 + 0.5\tanh(t-0.5)$, $\alpha_{2,4}(t) = 0.3 + 0.3\cos(2\pi 4t)$, $e_{2,2} = 2.002$, $e_{2,3} = 3.002$, and $e_{2,4} = 3.998$. The decomposition was performed in a deflationary manner by a two-step procedure in discrete time. To do this, we first solved the following least-squares problem

\begin{equation}
    \label{eq:Comp1}
    \hat{\gammab}_1 =
   \underset{\gammab_1\in \mathbb{R}^H}{\min}\ \left\|\tilde{\mathbf{x}}(n)- \left[\sum_{\ell=1}^{r_1} \alphab_{1,\ell}(n)\odot \boldsymbol{\Theta}_{1,\ell}(n)\right]\right\|^2,
\end{equation} 

\noindent with $\tilde{\mathbf{x}}(n) = \mathbf{x}(n)/\hat{\mathbf{B}}_{1,1}(n)$ and $\boldsymbol{\Theta}_{1,\ell}(n) = \cos\left(2\pi e_{1,\ell}\hat{\phib}_{1,1}(n)\right)+c_{1,\ell}\sin\left(2\pi e_{1,\ell}\hat{\phib}_{1,1}(n)\right)$. Estimates $\hat{\mathbf{B}}_{1,1}(n)$ and $\hat{\phib}_{1,1}(n)$ of the first component are obtained from the STFT of $\mathbf{x}$. The optimal coefficient vector $\hat{\gammab}_1$ was used to reconstruct the first component $\hat{\mathbf{x}}_1(n)$. Then, we computed the residual signal $\mathbf{x}_r(n) = \mathbf{x}(n)-\hat{\mathbf{x}}_1(n)$. Given a accurate estimation of the first component, then $\mathbf{x}_r(n)\approx \mathbf{x}_2(n)$ and the optimal coefficients for the second component can be obtained by solving 

\begin{equation}
    \label{eq:Comp2}
    \hat{\boldsymbol{\gamma}}_2 = 
   \underset{\gammab_1\in \mathbb{R}^H}{\min}\ \left\|\tilde{\mathbf{x}}_r(n)- \left[\sum_{\ell=1}^{r_2} \alphab_{2,\ell}(n)\odot \boldsymbol{\Theta}_{2,\ell}(n)\right]\right\|^2,
\end{equation}

\noindent where $\tilde{\mathbf{x}}_r(n) = \mathbf{x}_r(n)/\hat{\mathbf{B}}_{2,1}(n)$ and $\boldsymbol{\Theta}_{2,\ell}(n) = \cos\left(2\pi e_\ell\hat{\phib}_{2,1}(n)\right)+c_{2,\ell}\sin\left(2\pi e_\ell \hat{\phib}_{2,1}(n)\right)$. Estimates $\hat{\mathbf{B}}_{2,1}(n)$ and $\hat{\phib}_{2,1}(n)$ of the second component are obtained from the STFT of $\mathbf{x}_r$. This decomposition scheme was applied to noisy multicomponent signals $x(t) + n(t)$, with $n(t)$ being zero-mean Gaussian noise at $5$ different $\operatorname{SNR}$ levels: $0$, $5$, $10$, $15$ and $20$ dB. For each noise level, $100$ realizations were generated and the output $\operatorname{SNR}$ for each component was computed. As a baseline, our method was compared with the LR and SAMD results, which are also able to perform the simultaneous denoising and decomposition of multicomponent oscillatory signals. Thresholding methods are not capable of performing the decomposition of the signals, so they were not considered in this experiment. The top three rows of Fig. \ref{fig:multicomp} show the decomposition results for $x_1(t)$ and $x_2(t)$ separately, and the superposition $\hat{x}_1(t) + \hat{x}_2(t)$, obtained at $10$ dB of input noise, compared to the multicomponent signal $x(t)$. Our decomposition algorithm recovers both components accurately even when the wave-shape varies with time and in the presence of noise.

The complete decomposition results are shown in the last row of Fig. \ref{fig:multicomp}, where we show $\operatorname{SNR}_{out}$ vs. $\operatorname{SNR}_{in}$ for both components. From these curves, we see that starting from $\operatorname{SNR}_{in} = 5$ dB, our method outperforms LR and SAMD in the task of component reconstruction for both components. We achieved the best performance using our method although the variance is somewhat higher for all input noise levels. Many factors affect the performance of the algorithm, from the AM and PM estimates, optimal order $r$ and the adaptive number of nodes. Further analysis needs to be carried out on this decomposition task to increase the overall performance of the algorithm. Nevertheless, the proposed method performs better than the other methods based on the ANH model for all noise levels and competes with the hard-thresholding of the STFT in the monocomponent case.

\subsection{Segmentation of Signals with Sharp Transitions}
\begin{figure}
    \centering
    \includegraphics[width=\columnwidth]{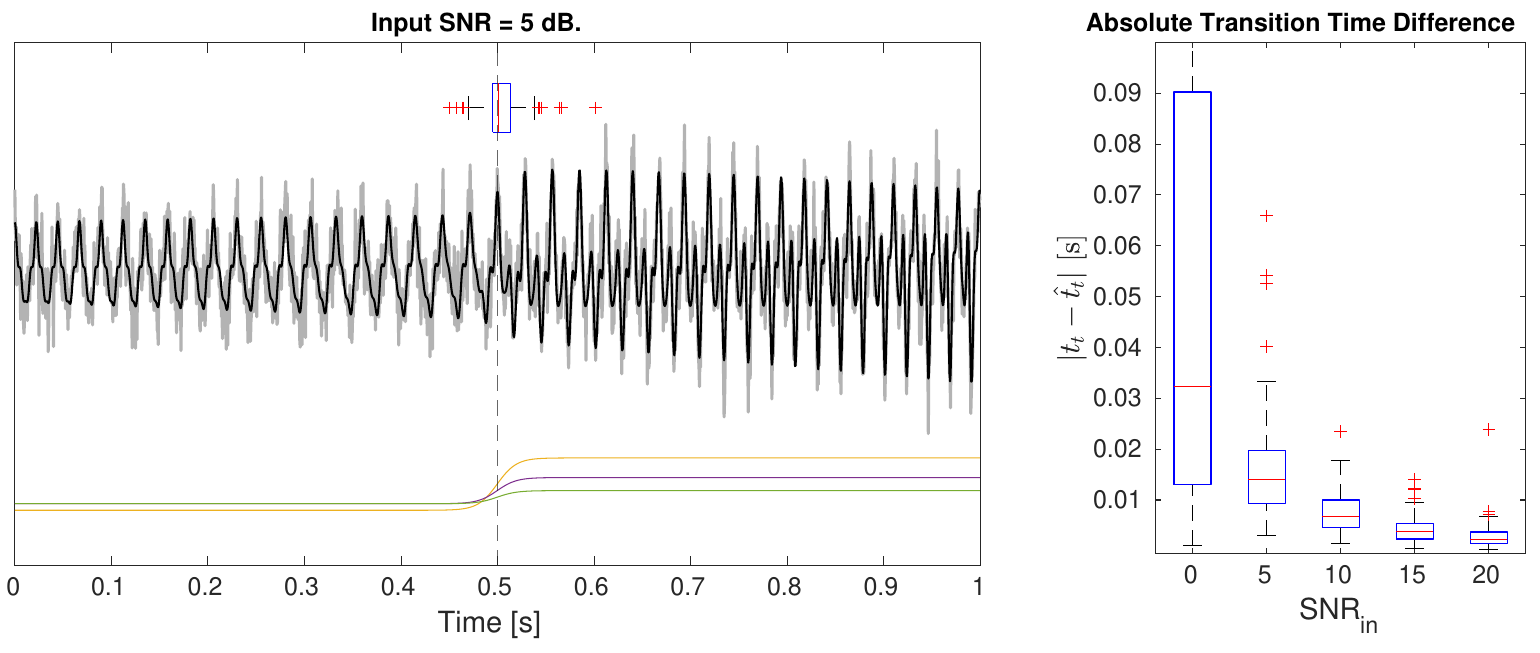}
    \caption{Left: Noisy non-stationary signal with sharp wave-shape transition plotted in gray. The noiseless signal is superimposed in black. The boxplot of the estimated transition times is plotted above and the theoretical HAFs are plotted below. Right: Boxplots of the absolute error (AE) between the real and estimated transition instants for different $\operatorname{SNR}_{in}$ levels.}
    \label{fig:segmentation}
\end{figure}

In this section, we consider the case of non-stationary signals with sharp transitions in their wave-shape. To model these transitions, we consider the following signal
\begin{equation}
    \label{eq:segment}
    x(t) = B_1(t)\cos(2\pi\phi_1(t)) + \sum_{\ell=1}^r B_{\ell}(t)\cos(2\pi e_\ell\phi_1(t)),
\end{equation}

\noindent where $B_1(t) = 0.1\sqrt{t+1}$, $\phi_1(t) = 40t + 5/(2\pi)\sin\left(2\pi t\right)$ and the HAFs $\alpha_\ell(t) = B_\ell(t)/B_1(t)$ are hyperbolic tangent functions $\mu + \lambda \tanh(\kappa(t-t_t))$, where $t_t$ is the transition time. For $t<t_t$, the modal amplitude would tend to $\mu - \lambda$, whereas for $t>t_t$ the modal amplitude would tend to $\mu + \lambda$. The parameter $\kappa$ controls the steepness of the transition at time $t_t$. This change in relative harmonic amplitude translates into abrupt changes in the wave-shape. In order to perform the segmentation of the signal, the time instants when the wave-shape significantly changes are to be detected. Based on the proposed model, the only transition happens at time $t_t$ so the problem reduces to locating this change point. We note here that of the methods discussed so far, only our proposal allows the detection of these abrupt wave-shape changes. Using our proposed wave-shape extraction procedure, the HAFs obtained by interpolation of the nodes $\alpha_{\ell,i}$ at locations $t_{\ell,i}$ codify the time-changing nature of the wave-shape. Given this, an adaptive transition time estimation procedure was developed based on the HAFs. For each HAF, the transition time is estimated using a change point detection algorithm~\cite{killick2012optimal} and the averaged time over all HAFs is reported as the estimated transition time $\hat{t}_t$. The left panel of Fig. \ref{fig:segmentation} shows an example signal with a sharp wave-shape transition at $t=0.5$. The signal is composed of $4$ harmonic components ($r=4$ in \eqref{eq:segment}). A noisy realization of the signal is shown in gray with the noiseless signal superimposed in black. A vertical line denotes the transition instant at $t_t = 0.5$. The first experiment comprised $100$ realizations of the noisy signal which were analyzed using our proposed method. For each, the transition point was estimated and the distribution of $\hat{t}_t$ is shown as a boxplot above the plotted signals. Additionally, the theoretic HAFs are shown below, demonstrating their effect on the wave-shape transition. We see from the boxplot that the method can accurately estimate the transition time even in the presence of noise. 

The robustness of the proposed segmentation algorithm was evaluated. To do this, $100$ signals with different sharp wave-shape transitions were simulated. In this experiment, the transition time $t_t$ is sampled from a uniform distribution in the interval $[0.1, 0.9]$. The transition parameter $\kappa$ is set to $50$. The amplitude parameters $\mu$ and $\lambda$ are sampled from uniform distributions in $[0.1, 0.5]$ and $[0.1, 0.35]$, respectively. The number of harmonic components $r$ is randomly chosen for each signal between $r = 3$ and $r = 6$. Zero-mean Gaussian noise was added to the signals at five input SNR levels: $0$, $5$, $10$, $15$ and $20$ dB. The signals were discretized at a sampling rate of $\Delta t = 0.5$ milliseconds. Our wave-shape extraction method was applied to each simulated signal and the transition time was estimated from the change point of the HAFs as described above. For each, the absolute error (AE) between the real transition time $t_t$ and our estimate was computed. The results of this experiment are shown in the right panel of Fig. \ref{fig:segmentation} using boxplots. Naturally, as the noise level decreases, the estimation error for the transition instant is reduced. We see that for $\operatorname{SNR}_{in}$ levels above $0$ dB most AE values fall below $20$ ms for a $1$ second duration signal. Moreover, for $\operatorname{SNR}_{in}> 10$ dB, the median value of the error distributions are in the order of magnitude of the sampling rate $\Delta t$. To the best of our knowledge, no other method exists that tackles the problem of detecting sharp transitions in the wave-shape of non-stationary signals.

\section{Applications in Real Signals}
\label{sec:real}
Real-world dynamical systems have a highly non-linear behavior that results in non-stationary signals. Particularly, physiological systems give rise to oscillatory biomedical signals that have time-varying amplitude, frequency and wave-shape. In this context, the adaptive non-harmonic model and its variations have been extensively applied to a variety of particular problems, including fetal and maternal ECG separation~\cite{su2017extract}, sleep apnea detection~\cite{lin2016sleep}, instantaneous heart rate estimation~\cite{lu2019recycling} and airflow recovery~\cite{huang2022airflow}, among others. To showcase the utility of our proposed algorithm, we analyze several biomedical signals including physiological and pathological cases. Firstly, we show a denoising example of the electroencephalogram (EEG) signal of a newborn during an epileptic seizure episode. Then, we perform the decomposition of an impedance pneumography signal (IP) into its respiratory and cardiac components. Finally, we use our algorithm to estimate the time-varying wave-shape of an electrocardiography signal (ECG) of a patient before and during ventricular fibrillation and use the information encoded in the second harmonic amplitude function to adaptively segment the signal. 

\subsection{Denoising of Epileptic Newborn Electroencephalography}
\begin{figure}
	\centering
	\includegraphics[width=\columnwidth]{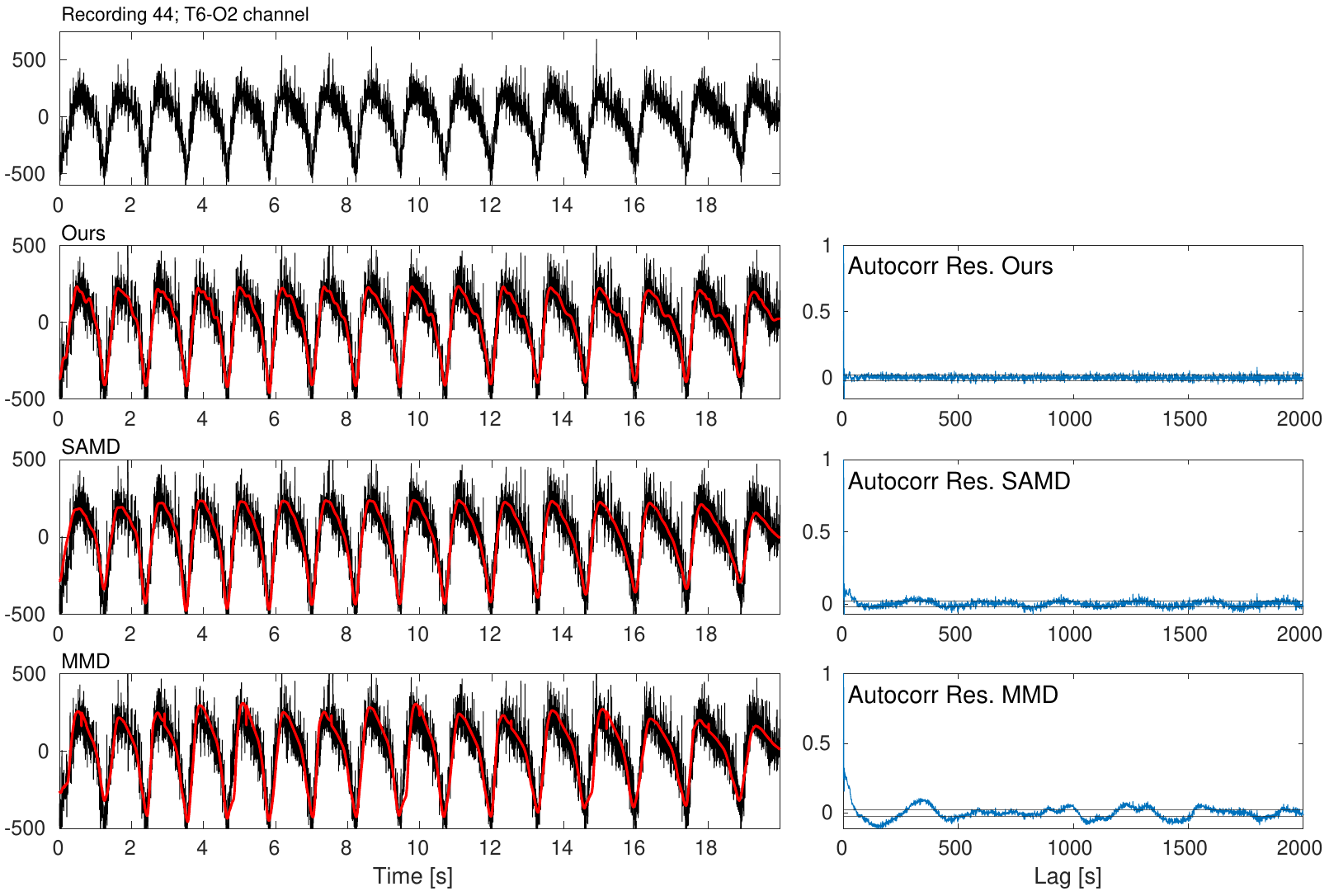}
	\caption{First Row: EEG signal during epileptic seizure event. Second row: EEG signal with our result superimposed (left) and ACF of the residual between the noisy and reconstructed signals (right). Third row: EEG signal with SAMD result superimposed (left) and ACF of the residual between the signals (right). Fourth row: EEG signal with MMD result superimposed (left) and ACF of the residual between the signals (right).}
	\label{fig:eeg}
\end{figure}

We analyzed an electroencephalography (EEG) signal of a newborn during an epileptic seizure episode. This signal is part of a publicly available dataset~\cite{stevenson2019dataset}. The EEG is shown in the top panel of Fig. \ref{fig:eeg}, where we clearly see the strong presence of noise originating from muscular contraction. We performed the denoising of the EEG signal using our algorithm and compared the resulting waveforms with those of SAMD and MMD. The signal was sampled at $256$ Hz and a window of $20$ seconds of the seizure event was considered for this analysis, resulting in a discrete signal with $5120$ samples. For the estimation of the fundamental amplitude and phase, the STFT was computed using a Gaussian window with $\sigma = 2\times 10^{-6}$, the maximum frequency jump for the ridge extraction algorithm was set to $I_f = 0.04$ Hz and the fundamental component reconstruction half-width was $\Delta = 0.4$ Hz. The adaptive WSF order obtained for this signal was $r = 6$. The resulting waveforms are shown in the first column of Fig. \ref{fig:eeg}. We see that the result of our method shows more variability between cycles of the signal when compared to the other methods. Particularly at the start and end of the signal, our proposal follows the seizure oscillation more accurately.  

In order to qualitatively compare the denoising results, the autocorrelation function (ACF) of the residuals was computed for each method. The normalized ACFs for our method, SAMD and MMD, respectively, are plotted on the right panels of Fig. \ref{fig:eeg}, alongside the $95\%$ confidence interval $\left[-1/\sqrt{N},1/\sqrt{N}\right]$. We see that our proposal leads to a residual with a structure closer to uncorrelated stationary noise. Additionally, the spectral entropy (SE) of the residual and the Pearson's correlation coefficient (PCC) between the residual and the output of the algorithms were computed. SE treats the normalized power spectrum of the residual as a probability distribution and its Shannon entropy is calculated. White uncorrelated noise is uniformly distributed in the frequency domain, leading to higher values of spectral entropy. The SE values for MMD, SAMD and our method were $6.71$, $7.10$ and $7.20$, respectively. As a reference, the SE value of uniform white noise with the same support is $7.34$. The SE of our result is closer to this value, showing the improved denoising obtained by our procedure. PCC is a normalized measure of the linear correlation between two data vectors. Values closer to $1$ ($-1$) indicate a strong positive (negative) correlation. We should expect that a better approximation of the signal results in less correlation (PCC closer to $0$) between the estimate and the residual error. MMD, SAMD and our method resulted in PCC values of $-0.25$, $-0.05$ and $-0.004$, respectively. These values show that our algorithm leads to a better estimation of the noiseless epileptic seizure waveform and better denoising performance than the other methods. 

\begin{figure}
	\centering
	\includegraphics[width=\columnwidth]{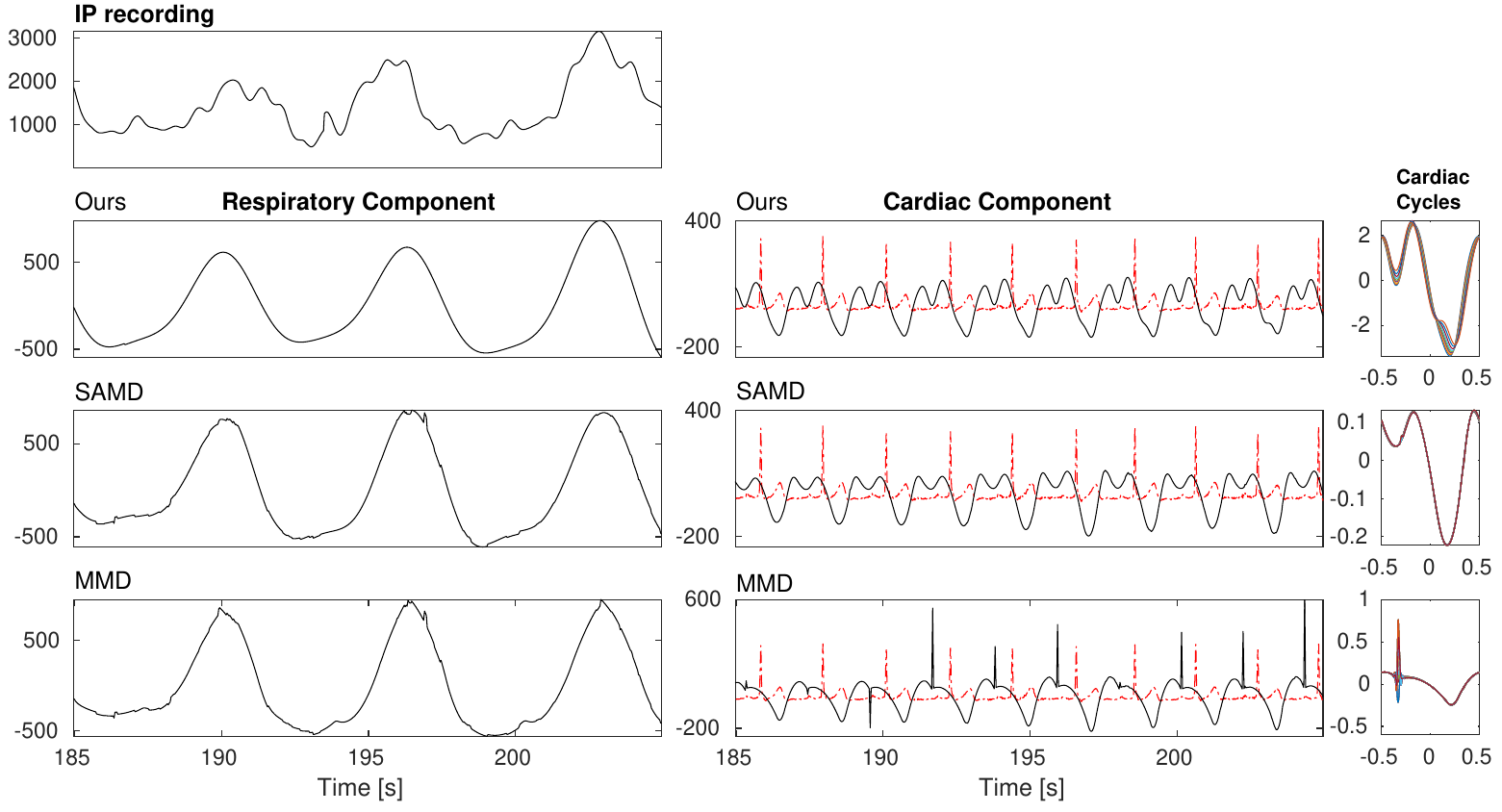}
	\caption{First row: $20$ seconds of the IP recording. Second row: respiratory and cardiac components extracted using our algorithm. Third row: respiratory and cardiac components extracted using SAMD. Fourth row: respiratory and cardiac components extracted using MMD. In each row, the ECG signal is plotted alongside the cardiac component with red dashed line. Segmented cardiac cycles for each method are shown in the right-most column.}
	\label{fig:ip}
\end{figure}
\subsection{Decomposition of Impedance Pneumography Signal}
In this section, we apply the wave-shape estimation algorithms to an impedance pneumography (IP) recording. This type of signal is normally composed of a respiratory component and a cardiac component, known as the cardiogenic artifact. This IP signal was recorded from a patient receiving flexible bronchoscopy examination using the Philips Patient Monitor MP60 at the Chang Gung Memorial Hospital, Linkou, New Taipei, Taiwan. The study protocol was approved by the Chang Gung Medical Foundation Institutional Review Board (No.104-0872C). The sampling frequency was $32$ Hz. SAMD, MMD and our algorithm were applied to $60$ seconds of the IP signal and the respiratory and cardiac components were extracted. The fundamental amplitude and phase of both components were estimated from the spectrogram using $\sigma = 10^{-6}$, $I_f = 0.3$ Hz and $\Delta = 0.008$ Hz. 
We show the resulting waveforms for the three decomposition methods in Fig. \ref{fig:ip}. With respect to the respiratory component, we see that the three methods eliminate the signal trend and are able to recover the waveform of the component. Our method results in a smoother waveform with slightly higher amplitude variability. This smoother waveform is more attractive to clinicians and allows better visualization of the information contained in the respiratory signal.

With respect to the cardiac component, both our method and SAMD recover a smooth time-varying wave-shape with each cycle corresponding to a heartbeat by comparing it to the simultaneously recorded ECG signal (plotted in red). Particularly, the wave-shape extracted using our proposal shows more cycle-to-cycle variability than SAMD, as can be seen in the segmented cardiac cycles in the right-most column of Fig. \ref{fig:ip}. The cardiac component extracted with MMD shows spiky artifacts that are not present in the original signal. Useful clinical information can be obtained from the cardiogenic artifact. In \cite{lu2019recycling}, the instantaneous heart rate (IHR) is estimated from the cardiogenic artifact extracted from the IP signal. In \cite{bucklar2003signal}, the stroke volume is estimated from the cardiac output curves obtained from the inductance cardiography signal after the removal of the respiratory component. The extracted cardiac component with time-varying wave-shape may be a useful alternative for cardiac output estimation.

\begin{figure}
	\centering
	\includegraphics[width=\columnwidth]{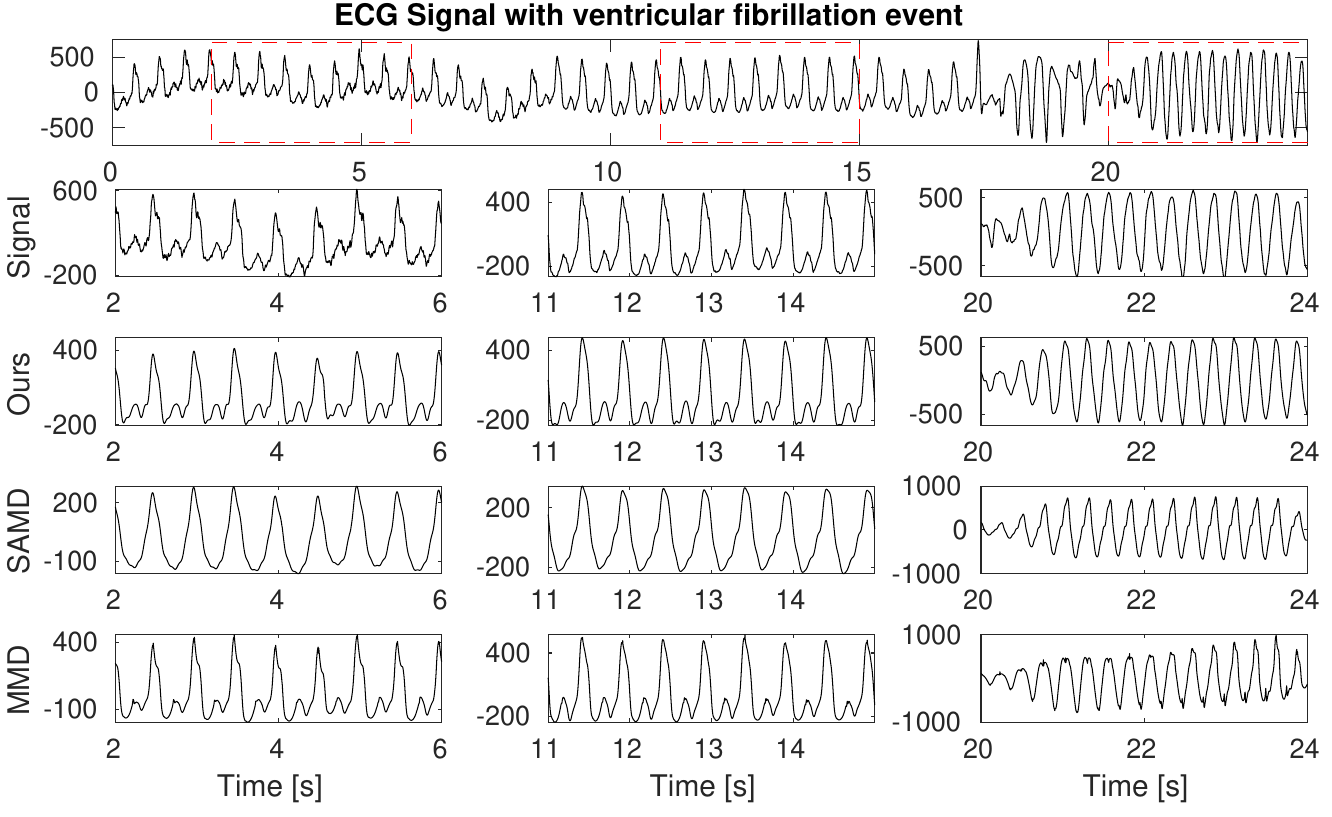}
	\caption{Top: ECG signal with ventricular fibrillation starting at $t = 17$ s. Illustrative segments at the start, middle and end of the signal are highlighted using red squares. Second row: $4$ second windows of the highlighted segments of the original signal. Third row: $4$ second windows of the highlighted segments of the time-varying wave-shape result using our method. Fourth row: same for the result using SAMD. Fifth row: same for the result using MMD.}
	\label{fig:ecg}
\end{figure}

\subsection{Segmentation of Electrocardiogram During a Ventricular Fibrillation Event}
In this section, we analyze an ECG of a tachyarrhythmic patient before and during a ventricular fibrillation (VF) event~\cite{nolle1986crei,goldberger2000physiobank}. VF is a pathological condition where the lower chambers of the heart contract rapidly in an uncoordinated fashion and leads to considerable changes in the waveform of the ECG. In this study, $24$ seconds of the ECG signal, sampled at $250$ Hz, were analyzed using our proposed algorithm, SAMD, and MMD. The fundamental component was estimated using $\sigma = 5\times 10^{-5}$, $I_f = 0.4$ Hz and $\Delta = 1.2$ Hz. The value of $r = 6$ was obtained using Wang criterion.

We show the ECG signal on the top panel of Fig. \ref{fig:ecg}. To illustrate the wave-shape extraction performance, segments at the start, middle and end of the original signal are highlighted with red squares. These same segments are shown in the second row of Fig. \ref{fig:ecg}. We show the concurrent segments of the reconstructed signal obtained with our method, SAMD and MMD in the third, fourth, and fifth rows, respectively. We see a clear change in the wave-shape between the first and second portions, associated with the rhythmic heartbeat, and the last portion, associated with the VF event. The estimated wave-shape using our method, SAMD and MMD are shown in the third to fifth rows, respectively. By analyzing the curves, we first note that all methods are able to perform \emph{detrending} and \emph{denoising} at the start of the signal, except for MMD which retains some of the trend. SAMD is unable to fully capture the time-varying wave-shape of the signal, particularly the more complex waveform associated with the normal heartbeat. When comparing our method to MMD, we see that both recover the normal beat wave-shape very accurately. When it comes to the VF waveform, MMD shows unnatural spiky artifacts at the peaks and valleys of the cycles alongside a slowly increasing trend. Moreover, in the first portion of the signal, MMD retains part of the original trend of the signal whereas our method is able to remove both the trend and the low amplitude noise.

\begin{figure}
	\centering
	\includegraphics[width=\columnwidth]{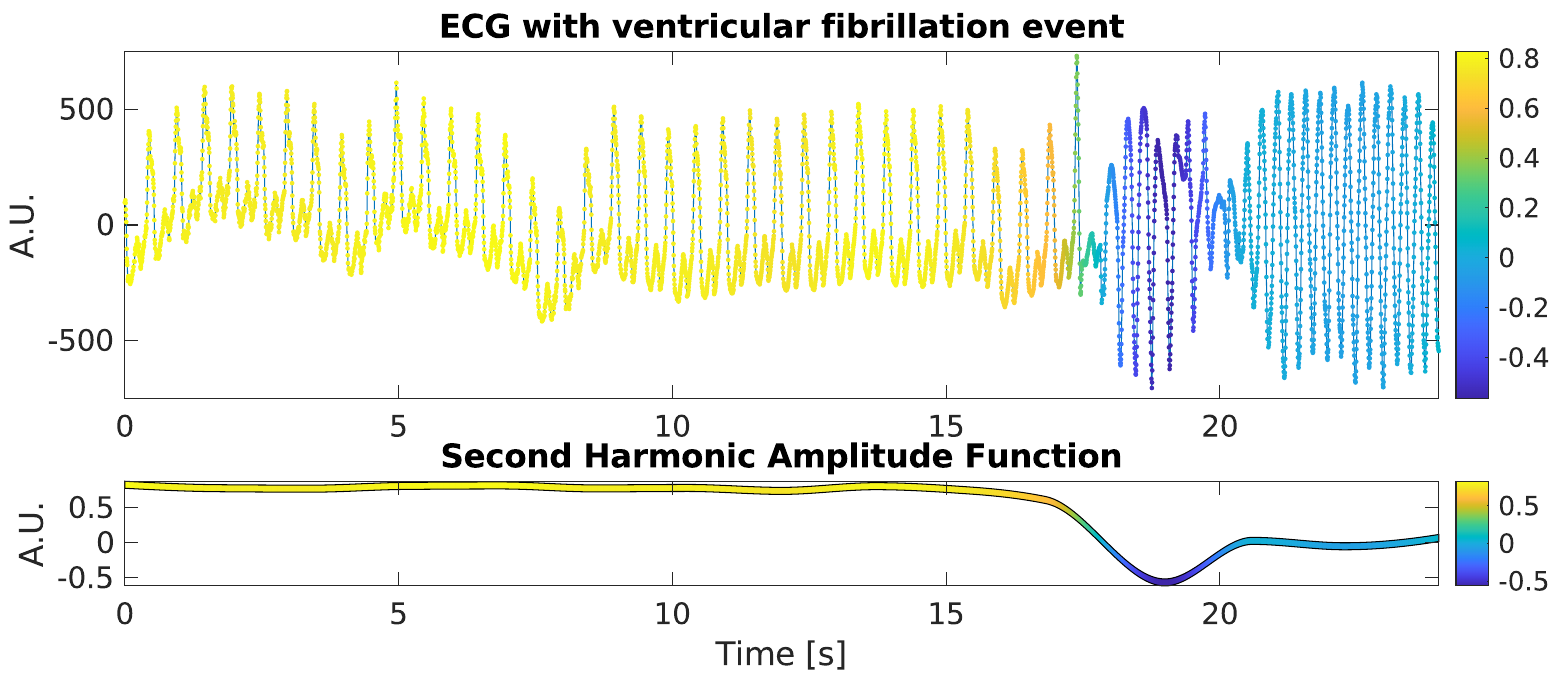}
	\caption{Top: Original ECG signal colored as function of $\alpha_2(t)$. Bottom: Second harmonic amplitude function $\alpha_2(t)$.}
	\label{fig:ecg_haf}
\end{figure}

Finally, we again show the ECG signal in the top panel of Fig. \ref{fig:ecg_haf} with the cycles colored as a function of the second harmonic amplitude function $\alpha_2(t)$. The latter is shown as a colored ribbon in the bottom panel of Fig. \ref{fig:ecg_haf}. We can see how different values of the harmonic amplitude are associated with the different wave-shapes of the signal. By comparing the HAF curve with the original signal shown on the top panel, we see a clear transition in the value of $\alpha_2(t)$ at the instant where the VF event starts. It is clear from this result that the temporal evolution of HAFs encodes the fast changes in the wave-shape and that significant events in the signal can be distinguished from the changes in the model parameters. This information can be useful for adaptive segmentation and event detection in biomedical signals under pathological conditions.

\section{Conclusions}
\label{sec:conclusions}
In this work, a novel time-varying wave-shape extraction algorithm based on the adaptive non-harmonic model was proposed for the analysis of non-stationary signals with time-varying wave-shape. The algorithm considers both relative harmonic phase and amplitude variations in order to characterize the variability in the wave-shape. The proposed method encodes the information associated with the time-varying harmonic amplitudes as a series of free nodes which can later be interpolated using shape-preserving cubic interpolators. Our method was validated using both synthetic and real-world signals for the tasks of denoising, decomposition and adaptive segmentation. We compared our method with existing algorithms for wave-shape estimation. In the case of synthetic signals, our method is able to characterize the time-varying nature of the wave-shapes much more accurately than the current methods for both monocomponent and multicomponent signals, even in the presence of high levels of noise. Similar results were obtained for real signals. For the denoising of the newborn epileptic EEG, our proposal was able to remove the noise without losing relevant information about the time-varying wave-shape of the signal, according to the measures considered. For the decomposition of the impedance pneumography signal, our method resulted in a smoother respiratory component that can be interpreted by clinicians more easily. With respect to the cardiac component, our method is able to recover a wave-shape with more variability than the other methods, which can be useful to obtain information about the underlying behavior of the circulatory system. For the ECG during ventricular fibrillation, our method was able to follow the abrupt changes in the wave-shape of the signal much more accurately than the currently available methods. Moreover, the coefficients associated with the second time-varying harmonic amplitude function were used to detect sharp transitions in the waveform of the ECG signal during ventricular fibrillation.

\section*{Acknowledgments}
This work was supported by the National Council for Scientific and Technological Research (CONICET) through the grant PIP-11220200100633CO; the National Agency for the Promotion of Technological and Scientific Research through the grants PICT-2020-SERIEA-01808, PICT-2020-SERIEA-01865; the National University of Entre Ríos (UNER) through the grants PID-UNER 6224, PID-UNER 6228; and Stic AmSud through the project STIC AmSud ASPMLM-Voice.

\section*{Conflict of interest}

The authors declare that they have no conflict of interest.

\bibliographystyle{elsarticle-num}
\bibliography{references}
   
\end{document}